\documentclass[11pt]{article}

\usepackage[a4paper,margin=2.55cm,includeheadfoot,headheight=14pt]{geometry}
\usepackage[T1]{fontenc}
\usepackage{amsmath,mathtools,bm}
\usepackage{newtxtext,newtxmath}
\usepackage{microtype}
\usepackage{booktabs}
\usepackage{array,tabularx}\usepackage{authblk}
\usepackage{graphicx}
\usepackage{placeins}
\usepackage{enumitem}
\usepackage{fancyhdr}
\usepackage{cite}
\usepackage{titlesec}
\usepackage[dvipsnames]{xcolor}
\usepackage[normalem]{ulem}
\definecolor{anyonblue}{HTML}{173B57}
\definecolor{anyonlink}{HTML}{245F88}

\numberwithin{equation}{section}
\usepackage[textsize=footnotesize]{todonotes}

\usepackage[colorlinks=true,linkcolor=anyonlink,citecolor=anyonlink,urlcolor=anyonlink]{hyperref}
\hypersetup{
  pdftitle={Interacting Anyons in a Harmonic Trap},
  pdfsubject={RG fixed points, many-body spectra, and the double-scaling phase diagram}
}

\titleformat{\section}
  {\Large\bfseries}{\thesection}{0.65em}{}
\titleformat{\subsection}
  {\large\bfseries}{\thesubsection}{0.6em}{}
\titleformat{\subsubsection}
  {\normalsize\bfseries}{\thesubsubsection}{0.55em}{}
\titlespacing*{\section}{0pt}{2.2ex plus 0.6ex}{0.9ex}
\titlespacing*{\subsection}{0pt}{1.8ex plus 0.5ex}{0.65ex}
\titlespacing*{\subsubsection}{0pt}{1.5ex plus 0.4ex}{0.5ex}

\setlist[itemize,enumerate]{topsep=0.5em,itemsep=0.35em,parsep=0pt}

\newcommand{\dd}{\mathrm d}
\newcommand{\ii}{\mathrm i}

\newcommand{\bA}{\bm A}
\newcommand{\bv}{\bm v}
\newcommand{\bx}{\bm x}
\newcommand{\bp}{\bm p}

\newcommand{\beq}{\begin{equation}}
\newcommand{\eeq}{\end{equation}}
\newcommand{\beqq}{\begin{equation*}}
\newcommand{\eeqq}{\end{equation*}}
\newcommand\beqa{\begin{eqnarray}}
\newcommand\eeqa{\end{eqnarray}}
\newcommand\beqaa{\begin{eqnarray*}}
\newcommand\eeqaa{\end{eqnarray*}}
\newcommand\bea{\begin{array}}
\newcommand\eea{\end{array}}
\newcommand{\la}[1]{\label{#1}}

\title{\vspace{-1.2cm}
\textbf{
Anyon Crystallization  by Statistics
 }}

\author[1]{Zohar~Komargodski}
\author[1]{Xuzixiang~Lou}
\author[2]{Ivri~Nagar}
\author[3,4]{Domenico~Orlando}
\author[4]{Susanne~Reffert}
\author[5]{Amit~Sever}
\affil[1]{Simons Center for Geometry and Physics, Stony Brook University, Stony Brook, NY, USA\\\href{mailto:zkomargo@gmail.com}{\nolinkurl{zkomargo@gmail.com}} \quad\href{mailto:xuzixiang.lou@stonybrook.edu}{\nolinkurl{xuzixiang.lou@stonybrook.edu}}}
\affil[2]{Center for Theoretical Physics --  a Leinweber Institute, Massachusetts Institute of Technology, Cambridge, MA, USA\\\href{mailto:ivri@mit.edu}{\nolinkurl{ivri@mit.edu}}}
\affil[3]{INFN, Sezione di Torino, Torino, Italy\\\href{mailto:domenico.orlando@to.infn.it}{\nolinkurl{domenico.orlando@to.infn.it}}}
\affil[4]{Albert Einstein Center for Fundamental Physics, Institute for Theoretical Physics, University of Bern, Bern, Switzerland\\\href{mailto:susanne.reffert@itp.unibe.ch}{\nolinkurl{sreffert@itp.unibe.ch}}}
\affil[5]{School of Physics and Astronomy, Tel Aviv University, Tel Aviv, Israel\\\href{mailto:asever@tauex.tau.ac.il}{\nolinkurl{asever@tauex.tau.ac.il}}}
\date{\today}

\begin{document}
\maketitle
\thispagestyle{plain}

\begin{abstract}

 We revisit the long-standing problem of the anyon gas. We study the solvable near-bosonic limit of a large number of anyons $N$ at fixed \(N/k\), where \(1/k\) characterizes the statistical interaction. In this limit, the ground state is a crystal with order $N/k$ lattice sites.  The statistical interactions, mediated by a long-range gauge field, are completely screened within the crystalline phase. The resulting state is a conventional crystal without superfluidity.
We characterize this anyon crystal by determining its unit-cell size, ground-state energy, acoustic excitations, angular momentum crossings, and other properties.

Our results indicate a quantum phase transition between the crystal at $k>k_c$ and a superfluid at $k<k_c$, with a critical value \(k_c\simeq2\text{--}5\). These results imply that an anyon gas with only statistical interactions is not necessarily a superfluid, contrary to common lore.

We also present a detailed analysis of the renormalization-group evolution of anyon--anyon interactions, identifying  many new potential fixed points and elucidating finite-size effects.

\end{abstract}

\newpage
\tableofcontents
\newpage
\section{Introduction}
\label{sec:introduction}

An anyon is a particle-like excitation in two spatial dimensions whose wave
function acquires a phase \(e^{i\theta}\) when two identical particles are
exchanged~\cite{LeinaasMyrheim,Wilczek1982}. %
We use the convention \(\theta=\pi/k\), so that \(k\to\infty\) corresponds to bosons and \(k=1\) to fermions. %
In nature, anyons arise as quasiparticles in fractional quantum Hall
states and, more broadly, in two-dimensional topologically ordered phases. They are interesting because their statistics are intrinsically two-dimensional, and these statistics can qualitatively
change the collective behavior of a many-particle system.

Already the problem of $N$ ``free'' anyons, governed by the Hamiltonian 
$H=\sum_{i=1}^N \frac{\bp_i^2}{2m}$,
is extremely nontrivial as the wave function $\Psi(\bx_1,\ldots,\bx_N)$ must account for the nontrivial statistics. If \(x_i\) and \(x_j\) are exchanged counterclockwise (with no other points inside the curve),  the wave function transforms as
\begin{equation}\label{exchangephase}
\bx_i \circlearrowleft \bx_j\ : \quad  \Psi\to e^{\pi\ii/k}\Psi\,.
\end{equation}  
Although point-like anyons exert no ordinary long-range force, their trajectories
are coupled through statistical Aharonov--Bohm phases: winding one particle
around another changes the many-body wave function by a  phase.  Determining the ground state in the  limit of many anyons remains a basic open question.

Much of the classic literature on anyons concerns their role as quasiparticles
of the fractional quantum Hall state.
The problem we consider here concerns a finite-density state of anyons. 
This is motivated by recent experiments on slightly doped fractional Chern insulators in twisted bilayer $\mathrm{MoTe}_2$, which have provided evidence for mobile fractionally charged excitations~\cite{XuSC2025,LiAnyonTrion2026}. These observations have motivated recent theoretical studies in which doping creates a finite-density gas of dispersive anyons~\cite{ShiSenthilPRX2025,ShiSenthilPNAS2025,Senthil2026}.

When the anyons are confined in a harmonic trap, 
\begin{equation}\label{Hharmonic}
	H=\sum_{i=1}^N \frac{\bp_i^2}{2m}+\frac{m\omega^2}{2} \sum_{i=1}^N \bx_i^2\,,
\end{equation}
the spectrum becomes discrete, making the spectral problem better defined.
The harmonic trap is particularly well-suited for the study of nonrelativistic (Schrödinger) conformal field theories.
Using the appropriate version of the state/operator correspondence, the energy of a state in the harmonic trap (in units of \(\omega\)) is identified with the conformal dimension of a corresponding operator~\cite{NishidaSon}. %

Some exact eigenstates of~\eqref{Hharmonic} are known: the ``chiral states'' whose energies depend linearly on the statistical
parameter $k^{-1}$.  These states are generally very far from the ground states but they are still interesting and we will review some of their properties. 

To ensure a self-adjoint Hamiltonian, we must specify the short-distance behavior of the wave function. In the limit where anyons are point-like, and barring any fine-tuning, the simplest choice is motivated by the existence of an infrared-stable fixed point of the renormalization group: we require that as  $\bx_i\to \bx_j$, the wave function decays as $|\bx_{ij}|^{\frac{1}{k}}$. %
Anyons can also be attractive rather than repulsive, but this case corresponds to infrared-unstable fixed points (\emph{i.e.} fine tuned fixed points) of the renormalization group. (One could regard it as the anyon analog of the Feshbach resonance, which is captured by the physics of fermions at unitarity~\cite{Feshbach:1958nx}.) We discuss these infrared-unstable fixed points and their required
multi-anyon short-distance data in Section~\ref{sec:contact} and Appendix~\ref{3anyonsapp}.

More generally speaking, however, the spectrum of interacting anyons is not known even for a few particles, and the thermodynamic problem, $N\to\infty$, is substantially harder. In this limit, is the system gapped? Or maybe a superfluid or a solid? This macroscopic characterization of the phase of many anyons should be largely independent of the trapping potential or confining geometry.

Close to the fermionic point, the anyon gas was analyzed extensively~\cite{Laughlin1988,FetterHannaLaughlin1989,Chen1989,LeeFisher1989,WenZee1990,BanksLykken1990,LykkenSonnenscheinWeiss1990,DaiEtAl1992,DuMehtaSon2021,GirardotRougerie2021}.
Available analytical and numerical
evidence supports a homogeneous superfluid phase.
The superfluid mode arises as the dual of the gapless Maxwell mode. Furthermore, the spectrum contains a roton-like minimum at finite
momentum.
Here we consider a different limit: near the bosonic point.
The
statistical interaction is weak in each elementary exchange but can act
coherently over a macroscopic number of particles.
In this coherent regime a semiclassical treatment becomes available.
Concretely, we will be studying the double-scaling limit
\begin{align}\label{eq:double-scalingintro}
	k\to\infty\,,\qquad N\to\infty\,,\qquad\text{with}\quad
    \lambda \equiv \frac{N}{k}\quad\text{fixed}.
\end{align}
After an appropriate field rescaling, \(1/k\) plays the role of an effective
Planck constant. In this paper, we concentrate on the large $k$ limit where the anyon problem is reduced to a controlled
semiclassical saddle-point problem.
We will argue that in this regime the system is a crystal.

\paragraph{Summary of Results.}

The second-quantized description is equivalent to the many-body Schr\"odinger problem~\eqref{Hharmonic}, but is more convenient for the large-$N$ 
limit. %
This description makes it clear that we are studying a repulsive, infrared-stable fixed point of nonrelativistic Chern--Simons matter theory.

In this description, the Chern--Simons term implements flux attachment via the Gauss law~\cite{JackiwPiClassicalQuantal,BergmanLozano}, and the two-body contact interaction fixes the anyon--anyon interaction to the infrared-stable fixed point. The action reads
\begin{equation}
 S=\int\dd t\,\dd^2x\left[
 \ii\Phi^\dagger D_t\Phi
 -\frac12(D_i\Phi)^\dagger D_i\Phi
 -\frac{\omega^2 r^2}{2}\Phi^\dagger\Phi
 -\frac{\pi}{k}\Phi^\dagger\Phi^\dagger\Phi\Phi
 +\frac{k}{4\pi}\epsilon^{\mu\nu\rho}
 A_\mu\partial_\nu A_\rho
 \right]\,,
 \label{eq:intro-CS-action}
\end{equation}
where $D_\mu=\partial_\mu+\mathrm{i}A_\mu$, and we use units in which $\hbar=m=1$. The Chern--Simons Gauss law attaches flux $2\pi/k$ to each $\Phi$ particle, thereby producing the anyonic
exchange phase $e^{\ii\pi/k}$. In addition, the Chern--Simons term modifies the angular momentum, as we will see.

It is useful to rescale the variables as 
\begin{align}
	\Phi=\sqrt{k}\,\phi\,,\qquad \phi=\sqrt{\rho}\,e^{-\ii\chi}\,.
 \label{eq:intro-density-phase}
\end{align}
The action (dropping total derivatives) then becomes proportional to \(k\),
\begin{equation}
S=k\int\dd t\,\dd^2x\left[
 \rho(\dot\chi-A_0)
 -\frac{1}{4\pi}\epsilon^{ij}A_i\dot A_j
 +\frac{A_0}{2\pi}B
 -\frac{(\nabla\rho)^2}{8\rho}
 -\frac{\rho}{2}(\nabla\chi-\bA)^2
 -\pi\rho^2
 -\frac{\omega^2 r^2}{2}\rho
 \right]\,. 
 \label{eq:intro-density-action}
\end{equation}
Thus, \(1/k\) is the effective
semiclassical expansion parameter, while the rescaled density obeys
\begin{equation}
    \int\dd^2x\,\rho=\frac{N}{k}\equiv\lambda\,,
\end{equation}
reflecting the fixed number of particles \(N\).
$A_0$ appears linearly, so we can remove it to obtain the Gauss law,
\begin{equation}
B\equiv\epsilon^{ij}\partial_iA_j =2\pi\rho\,.
 \label{eq:intro-gauss-law}
\end{equation}
Clearly, we obtain a semiclassical limit if we work in the double-scaling limit~\eqref{eq:double-scalingintro}.
In this limit, both the action and the particle-number constraint are of order one and the quantum fluctuations over the classical saddles are controlled by \(1/k\), which is our effective \(\hbar\).
The second-quantized description makes it manifest that~\eqref{eq:double-scalingintro} is a semiclassical limit.

Strictly speaking, the thermodynamic limit requires fixing $k$ while taking $N\to\infty$. However, in many similar double-scaling limits in other contexts~\cite{AGORLargeN,HellermanKrichevskiyOrlandoEtAl2024,DondiResurgence,SinghLargeCharge,BadelSemiclassics,GrassiKomargodskiTizzano,HellermanOrlandoSQCD,HellermanExponential}, the $\lambda\to\infty$ limit captures the physics and connects smoothly to the phase diagram at finite but large $k$. We will therefore solve the strong coupling limit $\lambda\to\infty$ and assume that it connects smoothly to the actual phase diagram in the thermodynamic limit. 

The same double-scaling limit~\eqref{eq:double-scalingintro}, in which the total statistical flux $\lambda$ remains finite, has been considered extensively in recent work, though not always at the infrared-stable fixed point of the renormalization-group flow of the anyon--anyon interaction~\cite{LundholmRougerie2015,CorreggiLundholmRougerie2017,LarsonLundholm2018,CorreggiDuboscqLundholmRougerie2019,Girardot2020,Nguyen2024,GirardotLee2024,AtaeiLundholmGirardot2025,Visconti2025,AtaeiEtAl2025}.

To find the ground state of~\eqref{eq:intro-density-action}, it is convenient to consider the corresponding Hamiltonian,
\begin{equation}
 \frac{H}{k}
 =\int\dd^2x\left[
 \frac{(\nabla\rho)^2}{8\rho}
 +\frac{\rho}{2}(\nabla\chi-\bA)^2
 +\pi\rho^2
 +\frac{ \omega^2 r^2}{2}\rho
 \right]\,,
 \qquad
 \nabla\times\bA=2\pi\rho\,,
 \qquad
 \int\dd^2x\,\rho=\lambda\,.
 \label{eq:intro-physical-Hamiltonian}
\end{equation}
Going beyond the classical limit, one must remember that, in this nonrelativistic system, $\chi$ and $\rho$ are conjugate variables and the components of $\bA$ are conjugate to each other. 

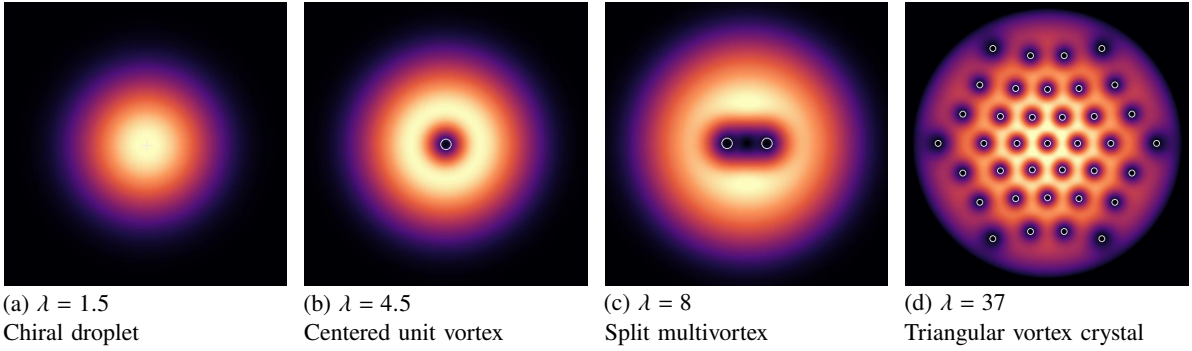
\begin{figure}[t]
  \centering
    \begin{tikzpicture}
    \def\panelw{0.235\textwidth}

    \node[inner sep=0pt, anchor=south west] at (0pt,0pt)
      {\includegraphics[width=\panelw]{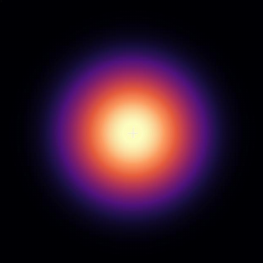}};
    \node[inner sep=0pt, anchor=south west] at (0.25\textwidth,0pt)
      {\includegraphics[width=\panelw]{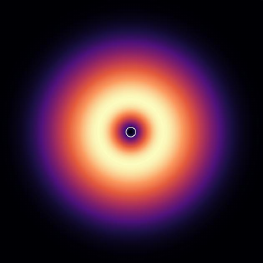}};
    \node[inner sep=0pt, anchor=south west] at (0.50\textwidth,0pt)
      {\includegraphics[width=\panelw]{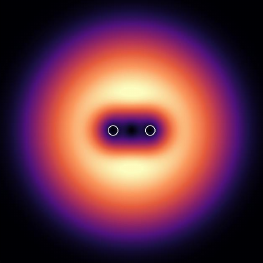}};
    \node[inner sep=0pt, anchor=south west] at (0.75\textwidth,0pt)
      {\includegraphics[width=\panelw]{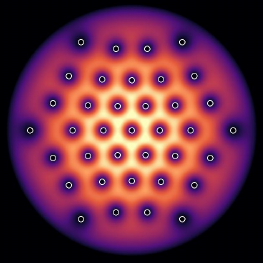}};

    \node[anchor=north, align=left, text width=\panelw,
          font=\footnotesize, inner sep=0pt]
      at (0.1175\textwidth,-3pt)
      {(a) $\lambda = 1.5$\\Chiral droplet};

    \node[anchor=north, align=left, text width=\panelw,
          font=\footnotesize, inner sep=0pt]
      at (0.3675\textwidth,-3pt)
      {(b) $\lambda = 4.5$\\Centered unit vortex};

    \node[anchor=north, align=left, text width=\panelw,
          font=\footnotesize, inner sep=0pt]
      at (0.6175\textwidth,-3pt)
      {(c) $\lambda = 8$\\Split multivortex};

    \node[anchor=north, align=left, text width=\panelw,
          font=\footnotesize, inner sep=0pt]
      at (0.8675\textwidth,-3pt)
      {(d) $\lambda = 37$\\Triangular vortex crystal};
  \end{tikzpicture}
  \caption{A schematic summary of the phases found in the double-scaling limit~\eqref{eq:double-scalingintro}: (a) The chiral zero-vortex solution at
  $\lambda=1.50$. 
  (This state is known exactly to all orders in the $1/k$ expansion, as reviewed in the main body of the paper.) (b) A centered unit-vortex
  solution. (c) The split
  multivortex state (centered
  multivortices are unstable due to vortex--vortex repulsion). (d) The large-$\lambda$ asymptotic
triangular vortex crystal. The crystal is screened in the sense that, after coarse-graining over several
unit cells, the vortex density approaches the matter density. The unscreened component is depleted in the bulk and survives predominantly near the edge at finite \(\lambda\).}
  \label{fig:ground-states-vs-lambda}
\end{figure}

Minimizing the energy~\eqref{eq:intro-physical-Hamiltonian} as a function of $\lambda$,
we find the sequence of classical saddles shown schematically in Figure~\ref{fig:ground-states-vs-lambda}.%
\begin{itemize}
\item For small values of \(\lambda\), the lowest energy state is the rotationally invariant ``chiral'' zero-vortex solution. By ``chiral'' we mean that the wave function has a purely holomorphic piece. This solution describes a simple density profile of the anyons. It is the semiclassical version of a known exact anyon state and describes a smooth droplet of matter.  In terms of the original Chern--Simons matter theory~\eqref{eq:intro-CS-action}, it is a standard Higgs phase. 
\item Around \(\lambda \approx 2.96\), a unit vortex appears at the center of the trap. This is an isolated zero of the density profile, around which there is vorticity in $\chi$.
\item At larger \(\lambda\) (starting from $\lambda\sim 5.1$), additional vortices appear in the droplet.  The centered vortices of winding \(\ell\geq2\) are unstable to splitting, so the relevant configurations contain separated elementary vortices and no longer preserve rotational invariance. At sufficiently large \(\lambda\), these vortices form a locally triangular lattice.\footnote{A triangular vortex lattice was also found
in~\cite{CorreggiDuboscqLundholmRougerie2019,AtaeiEtAl2025} by numerically minimizing the almost-bosonic average-field energy functional.}
\item The \(\lambda\to\infty\) asymptotic limit is a crystal state with a triangular vortex lattice. It has an important property of being completely screened, \emph{i.e.}, on scales large compared with the vortex spacing, the vortex circulation cancels the average statistical magnetic field. That means that the matter density is proportional to the density of vortices and the long-distance statistical interaction is screened in the crystalline state.

For this reason, the large-\(\lambda\) phase is best regarded as an ordinary
crystal of anyonic composites.  It is not a supersolid.  In particular, there
is no independent superfluid Goldstone mode: a smooth phase rotation can be
absorbed into a flat statistical gauge connection. Its gapless
long-wavelength degrees of freedom are instead the longitudinal and
transverse acoustic phonons of the triangular lattice. Contrary to common lore, the crystal is stabilized solely by the statistical interactions in~(\ref{exchangephase}), without any Coulomb forces.

\end{itemize}
  
To reiterate the key point, the crystalline state at $\lambda\to\infty$ is completely screened: the average anyon density, coarse-grained over several unit cells, coincides with the density of vortices
\begin{equation}
	\frac{\rho_v}{\bar\rho}=1\,,\qquad(\lambda\to\infty)\,.
\end{equation}
(Here and later in the paper we normalize the density of anyons by a factor of $1/k$ so that a completely screened state corresponds to $\frac{\rho_v}{\bar\rho}=1$.)  
Therefore $k$ fundamental anyons accompany the vortices.  This makes it an ordinary vortex crystal rather than a supersolid. %
It is completely screened because $k$ anyons collectively are transparent, as their magnetic field is a multiple of $2\pi$.

For large \(\lambda\), we are effectively in the local-density approximation and the coarse-grained charge density takes a Thomas--Fermi form
\begin{equation}\label{TFintro}
   \bar \rho(r) =\frac1{2\widehat e(1)}
 \left[\sqrt{\frac{2\widehat e(1)\lambda}{\pi}}
       -\frac{r^2}{2}\right]\,,
\end{equation}
for \(r<R=(8\widehat e(1)\lambda/\pi)^{1/4}\), with \(\bar\rho(r)=0\) outside. Solving the periodic unit-cell problem, we find that the triangular lattice
is energetically preferred and obtain
\begin{equation}
    \widehat e(1)=7.223\pm0.004\,.
\end{equation}
This gives the leading large-\(\lambda\) energy and radius in the harmonic
trap,
\begin{equation}
    \frac{E_0}{\omega}
    = \frac{4}{3} \sqrt{\frac{\widehat e(1)}{2 \pi k}} N^{3/2} = 1.43\frac{N^{3/2}}{\sqrt{k}}\,,
    \qquad
    R=2.07\left(\frac{N}{k}\right)^{1/4}\,.
\label{eq:crystal-scaling-summary}
\end{equation}
The \(N^{3/2}\) behavior is the expected large-charge scaling for a
Schr\"odinger-invariant theory in a harmonic trap~\cite{FavrodOrlandoReffert2018,KravecPal2019}. This shows that our double-scaling limit indeed captures macroscopic phases of anyon matter.

The system is expected to form a crystal essentially in any strong enough trap.
  Since our treatment is limited to the leading bulk terms, ignoring for example boundary effects, we can easily generalize it to other geometries, including the hard trap of radius \(R\).
  Conformal invariance fixes the form of the equation of state: the energy density must be proportional to the square of the charge density.
The proportionality factor is nothing else than the constant \(\widehat e(1)\) appearing in the Thomas--Fermi distribution in Eq.~\eqref{TFintro}:
\begin{equation}\label{eq:eosnum}
\varepsilon(\rho)=\frac{\widehat e(1)}{k}\rho^{2}~.
  \end{equation}
  It follows that the energy in the hard trap is given by
  \begin{equation}
    E_{\text{hard trap}} = \frac{1}{\pi R^2} \frac{\widehat e(1)}{k} N^2~.    
  \end{equation}

Extrapolating this behavior to the thermodynamic limit, we can compare the energy in~\eqref{eq:crystal-scaling-summary} with the energy of the translation-invariant superfluid state to estimate where the crystal turns into a superfluid.
This estimate gives $k_c\sim2\text{--}3$. %

The zero-point motion of the vortex lattice, associated with the phonons, provides another method to estimate the melting point.
Using the elastic moduli of the crystal,
\begin{equation}
    K=14.446\pm0.008\,,
    \qquad
    G=1.354\pm0.004\,,
\end{equation}
we determine the two acoustic modes,
\begin{equation}
    \omega_L(q)=\sqrt{\bar\rho(K+G)}\,q+\cdots\,,
    \qquad
    \omega_T(q)=\sqrt{\bar\rho G}\,q+\cdots\,.
\end{equation}
From these we can apply the usual Lindemann criterion to find
\begin{equation}
\frac{\sqrt{\langle|\bm u|^2\rangle}}{a_v}
\sim
\frac{0.52}{\sqrt{k}},
\end{equation}
where \(a_v\) is the nearest-neighbor vortex spacing and \(\bm u\) is the displacement field of the vortex crystal. This is compared with the standard Lindemann threshold, for which the ratio is typically \(\simeq0.23\text{--}0.28\), corresponding to \(k_c\simeq3.5\text{--}5\).
These estimates are roughly consistent, suggesting that the transition occurs around
\begin{equation}\label{estintro}
k_c\sim 2\text{--}5.
\end{equation}
We will mention yet another estimate in the bulk of the paper.

In summary, we suggest the following simple phase diagram for the anyon gas. For \(k>k_c\), the system is in a crystalline state, while for \(k<k_c\) it is in a superfluid state.\footnote{We do not gauge $U(1)_{\mathrm{EM}}$; if gauged, the superfluid is a superconductor and the crystal is an ordinary insulator.} At \(k=k_c\), a first-order phase transition is expected.\footnote{\label{prelimMonte} We have attempted preliminary variational Monte Carlo calculations and seen evidence for a transition between a solid and superfluid between $k_c=2-3$, consistent with~\eqref{estintro}. Furthermore, 
variational Monte Carlo gives
\begin{equation}
\widehat{e}(1)=6.85 \pm 0.31~,
\end{equation} beautifully consistent with the calculations of the equation of state in the double scaling limit~\eqref{eq:eosnum}. We hope to return to this in the future for a systematic treatment. }

There may already be partial evidence for this scenario.
The homogeneous superfluid near the fermionic endpoint, \(0<k-1\ll1\), has pronounced roton-like excitations~\cite{DuMehtaSon2021}. In addition, numerical calculations at the semion point, \(k=2\), find a linearly dispersing collective mode together with a roton minimum at finite momentum~\cite{XieHeDasSarma1990}. The roton need not close its gap for the solid state to become energetically favored, since cubic terms can lower the energy of the crystalline state even before the roton gap closes~\cite{PomeauRica1994}.

Finally, we analyze the renormalization group of short-range anyon
interactions.  The repulsive two-body fixed point used above is
infrared-stable.  The attractive fixed point is more subtle: at fixed \(k\),
multi-anyon contact operators become relevant in sufficiently large particle
number sectors, so the two-body boundary condition alone does not define a
complete nonrelativistic many-body fixed point.  We discuss this issue explicitly in the
two- and three-anyon sectors and describe the resulting
structure.

\paragraph{Outline.}

In Section~\ref{sec:contact}, we consider the renormalization group analysis of anyon--anyon interactions. We present the $N=2$ anyon problem and the exact wave functions, beta functions, and scaling dimensions. In Section~\ref{sec:double-scaling}, we consider the double-scaling limit and derive the results quoted above. In Section~\ref{sec:melting}, we extrapolate beyond the double-scaling limit and explain the phase diagram of the anyon gas. In Appendices~\ref{app:numerical_methods}--\ref{sec:crossing-density}, we summarize some of the numerical methods used in this work,  present numerical evidence for the approach to complete screening at large $\lambda$, further analyze the unstable fixed point and its multi-anyon spectrum, derive the large-$\lambda$ giant-vortex asymptotics and show that a macroscopic giant vortex undergoes fission, and finally determine the density of angular-momentum crossings.

\section{Aspects of the Theory of Anyons}
\label{sec:contact}

\subsection{Renormalization Group Analysis of Anyon Interactions}

For the renormalization group analysis, we %
consider the quartic translation-invariant action 
\begin{equation}
 S=\int\dd t\,\dd^2x\left[
 \ii\Phi^\dagger D_t\Phi
 -\frac12(D_i\Phi)^\dagger D_i\Phi
-\gamma\,\Phi^\dagger\Phi^\dagger\Phi\Phi
 +\frac{k}{4\pi}\epsilon^{\mu\nu\rho}
 A_\mu\partial_\nu A_\rho
 \right]\,,
 \label{eq:sec2-CS-action}
\end{equation}
where we have dropped the harmonic trap because it is not relevant for the short-distance physics. The parameter $\gamma$ captures s-wave two-body interactions.
In principle, we should also consider three-body interactions, interactions among different angular momentum modes, and so on.

We will now analyze these issues and, in particular, explain why we used $\gamma=\pi/k$ in the introduction~\eqref{eq:intro-CS-action}. 

In nonrelativistic theories, beta functions and scaling dimensions can often be calculated exactly.
This is because there is no particle creation and thus it suffices to solve the quantum mechanics of a fixed number of particles. 
In particular, to understand the exact renormalization group flow of $\gamma$ it is enough to solve the problem of two anyons.

\subsubsection{The Two-Anyon Problem}

We begin with the two-anyon sector, supplementing the $N=2$ Hamiltonian with the contact interaction
\begin{equation}
    V(\bm r)=g\,\delta^{(2)}(\bm r)\,,
    \label{eq:physical-contact}
\end{equation}
where $\bm r=\bm r_1-\bm r_2$ is the relative coordinate. The coupling $g$ is related to $\gamma$ in (\ref{eq:sec2-CS-action}) by $\gamma=\mu g$, with $\mu$ the reduced mass. Although this interaction is classically scale invariant in two spatial dimensions, quantum effects induce a renormalization-group flow for $g$. We now analyze this flow and identify its fixed points.

We impose the exchange statistics on the two anyon wave function
\begin{align} \label{eq:exchange}
\psi(r,\varphi+\pi)=e^{\ii\theta}\psi(r,\varphi),\qquad  0\leq\theta<\pi\,.
\end{align}
The allowed relative angular momenta are therefore
\begin{align}
	 \ell=2n+\theta/\pi,\qquad n\in\mathbb Z.
  \label{eq:angular-momenta}
\end{align}
(The interval $\pi<\theta\leq 2\pi$ is related to $0\leq\theta<\pi$ via time reversal. We will comment below on how this arises. See also Appendix~\ref{3anyonsapp}.) 
The limit $\theta=0$ corresponds to standard bosonic statistics, while
$\theta=\pi$ is fermionic.

Writing
\begin{equation}
  \psi(r,\varphi)=\frac{1}{\sqrt r}\,e^{\ii\ell\varphi}u_\ell(r),
\end{equation}
the radial equation is
\begin{equation}
  \left[-\frac{1}{2\mu}\frac{\dd^2}{\dd r^2}
  +\frac{\ell^2-\frac14}{2\mu r^2}+g\,\delta^{(2)}(\bm r)\right]u_\ell(r)
  =E u_\ell(r)\,.
  \label{eq:radial-Schrödinger}
\end{equation}
The delta-function term in this equation is naively dimensionless and the system is naively scale invariant. However, there is a nontrivial renormalization group equation for $g$, which develops an anomalous dimension. 

Near $r=0$, the two independent behaviors are
\begin{equation}\label{smallr}
	 u_\ell(r)\sim r^{\lambda_\pm}\,,\qquad
\lambda_\pm=\frac12\pm|\ell|\,.
\end{equation}
Because the norm is proportional to $\int_0^\infty\dd r\,|u_\ell(r)|^2$,
the irregular solution is locally normalizable only when
$|\ell|<1$. For $0<\theta<\pi$, the only channel satisfying this condition is
$n=0$, for which $\ell=\theta/\pi$.

To regularize the delta function, we
excise a disk of radius $R$ around the coincidence point and define the regulated theory on $r\geq R$. We then add a localized potential term to the Hamiltonian on the circle of radius $R$,
$H=\cdots+g(R)\delta_R^{(2)}(\bm r)$, where
\begin{equation}
\delta_R^{(2)}(\bm r)
      \equiv \frac{\delta(r-R)}{2\pi R}\,.
\end{equation}
Finally, to ensure that this Hamiltonian is self-adjoint in the domain $r\geq R$, we must impose~\cite{AmelinoBak} 
\begin{equation}
R\frac{\partial_r\psi(R)}{\psi(R)}=\frac{\mu g(R)}{\pi}\,.
\end{equation}
At very low energies, the solution in the exterior takes the form
\begin{equation}
    \psi(r,\varphi)
    =
    e^{\ii{\theta\over\pi}\varphi}
    \left(
        c_+ r^{\theta\over\pi}+c_- r^{-{\theta\over\pi}}
    \right).
\end{equation}
It follows that
\begin{equation}
    g(R)
    =
    \frac{\theta}{\mu}
    \frac{
        c_+R^{\theta\over\pi}-c_-R^{-{\theta\over\pi}}
    }{
        c_+R^{\theta\over\pi}+c_-R^{-{\theta\over\pi}}
    }\,.
\end{equation}
Holding the physical ratio $c_-/c_+$ fixed (this ratio is related to the asymptotic phase shift, for instance) and differentiating with
respect to the cutoff gives
\begin{equation}\label{eq:beta-g}
    \beta_g
    \equiv
    -R\frac{\dd g}{\dd R}=\frac{\mu}{\pi}
        \left(
            g-\frac{\theta}{\mu}
        \right)
        \left(
            g+\frac{\theta}{\mu}
        \right)\,.
\end{equation}

The two fixed points are thus
\begin{equation}\label{gfixedpoints}
g_\pm=\pm\frac{\theta}{\mu}\,.%
\end{equation}
In the near-field limit, they select,
respectively, $\psi(r,\varphi)\propto e^{\ii{\theta\over\pi}\varphi}r^{\pm{\theta\over\pi}}$. 
The plus sign is the regular, repulsive, infrared-stable fixed point, whereas
the minus sign is the irregular, attractive, infrared-unstable fixed point.

Having understood the fixed points, we can now take $R\to 0$ and return to our original problem~\eqref{eq:radial-Schrödinger}.
In the conventions of the field-theory action~\eqref{eq:sec2-CS-action}, the particle
mass is one, so $\mu=1/2$, and the quartic interaction produces the two-body
potential $g\delta^{(2)}(\bm r)=2\gamma\delta^{(2)}(\bm r)$.
Consequently $\gamma=\mu g$, and the fixed points are
\begin{equation}
\gamma_\pm=\pm\theta=\pm\frac{\pi}{k}\,.
\end{equation}
This explains why we choose $\gamma=\pi/k$ in the rest of the paper: it is the infrared fixed point of repulsive point-like anyons. 

The slope of the beta function at the fixed point gives the scaling dimension of the quartic term,
\begin{equation}
\Delta[\Phi^\dagger \Phi^\dagger \Phi\Phi]=4+\frac{2}{k}\,.\label{eq:dimensions}
\end{equation}
This result is exact.
Taking into account the anyons' finite size $R$ in a system of size \(L\), we should expect corrections of order $(R/L)^{2/k}$.
We ignore these effects here and focus on anyons at the infrared-stable fixed point.
As a consistency check of the exact result~\eqref{eq:dimensions}, one can set $k=1$, which yields $\Delta=6$. Indeed, the exclusion principle forbids quartic fermion interactions, and the first interaction is sextic.

Another way to derive~\eqref{eq:dimensions} is to calculate the exact two-anyon spectrum in a harmonic trap and then use the operator-state correspondence.
The spectrum is obtained as follows.
After separating the center of mass, the relative Hamiltonian is the same as in~\eqref{eq:radial-Schrödinger} with 
$\mu=1/2$ and an additional radial potential $\frac{\mu\omega^2r^2}{2}$.
Its eigenfunctions are written in terms of generalized Laguerre polynomials,
\begin{equation}
  u_{n_r,\ell}(r)
  \propto
  r^{\frac12+|\ell|}
  e^{-\mu\omega r^2/2}
  L_{n_r}^{|\ell|}(\mu\omega r^2),
\end{equation}
with eigenvalues
\begin{equation}
  E_{\rm rel}
  =\omega\left(1+2n_r+|\ell|\right),
  \qquad n_r=0,1,2,\ldots .
\end{equation}
Adding the center-of-mass zero-point energy gives
\begin{equation}
  \frac{E}{\omega}
  =2+2n_r+|2n+1/k|\,.
\label{exacttwobody}\end{equation}
For $n=0$ we obtain $\frac{E}{\omega}=2+1/k+2n_r$, from which we find the scaling dimension
\begin{equation}
    \Delta[\Phi^\dagger\Phi^\dagger]    =2+\frac{1}{k}\,.
\end{equation}
However, in nonrelativistic theories, operators with positive and negative particle number factorize~\cite{BoisvertFaddaKulpYazdi}, and~\eqref{eq:dimensions} follows.

So far we have described the interval $\theta\in[0,\pi)$. (Note that at \(\theta=\pi\), the irregular solution in \eqref{smallr} has a logarithmically divergent norm, leaving a unique admissible boundary condition.)
For $\theta\in (\pi,2\pi]$, the $n=-1$ modes require a similar self-adjoint completion. Therefore $\theta\in [0,\pi)$ and $\theta\in (\pi,2\pi]$ lead to an isomorphic spectrum in a slightly nontrivial way, see Appendix \ref{3anyonsapp} for more details. 

\subsubsection{Comments about the Unstable Fixed Point}

Let us now comment briefly on the unstable fixed point. 
At the unstable fixed point we obtain, analogously to~\eqref{eq:dimensions}, $\Delta[\Phi^\dagger \Phi^\dagger \Phi\Phi]=4-\frac{2}{k}$.
This by itself does not present a paradox.
However, as we will review below, considering $N$ particles in the harmonic trap reveals states that are dual to the operators $(\Phi^\dagger)^N$ with scaling dimension $\Delta=N-\frac{1}{2k}N(N-1)$.
Therefore operators such as $(\Phi^\dagger)^N\Phi^N$
become relevant and eventually violate the unitarity bound at any given fixed $k$.
Therefore, at any fixed $k$, the renormalization group analysis above of the coupling constant $\gamma$~\eqref{eq:beta-g} is incomplete and does not describe a unitary nonrelativistic fixed point with $\gamma=-\pi/k$.

To determine if a true, unitary, multi-critical fixed point exists, one must attempt to deform the action by these $N$-body operators and search for a fixed point.
This corresponds to modifying the boundary conditions for when $N$ anyons meet simultaneously.
In other words, the rule that the wave function blows up as $r^{-{\theta\over\pi}}$ when two anyons meet does not by itself lead to a complete, sensible theory, since one must specify what happens when more than two anyons approach each other.

Therefore, it remains an open question whether the fixed point $\gamma=-\pi/k$ exists.
Nonetheless, much of the literature is concerned with $\gamma=-\pi/k$ since it has interesting classical solutions~\cite{JackiwPiSolitons,JackiwPiClassicalQuantal,JackiwPiReview,JackiwPiTimeDependent}. In Appendix~\ref{3anyonsapp}, we analyze the sectors containing more than two anyons at the unstable fixed point. We provide pieces of evidence for the existence of a rich family of fixed points distinguished by their three- and higher-anyon contact interactions. In addition, we show that the dimensions of the chiral operators remain above the unitarity bound at these new fixed points that we identify.

\subsection{Some Exact Solutions of \texorpdfstring{$N$}{N} Anyons in a Harmonic Trap}\label{chiralsec}

Quite remarkably, some exact wave functions are known for the $N$-anyon system in a harmonic trap. 
Let $z_i=(\bx_i)_1+\ii (\bx_i)_2$. The simplest exact anyon wave functions are\footnote{Henceforth we set \(\omega=1\).}
\begin{equation}
\Psi_{N,+}\propto
 e^{-\frac12\sum_i|z_i|^2}\prod_{i<j}(z_i-z_j)^{1/k}\,,
 \qquad
 \Psi_{N,-}\propto
 e^{-\frac12\sum_i|z_i|^2}\prod_{i<j}(\bar z_i-\bar z_j)^{-\frac{1}{k}}\,,
\label{eq:chiral-anyon-gauge}
\end{equation}
where $+$ and $-$ correspond to the infrared-stable and infrared-unstable fixed points, respectively.
These are eigenfunctions of $H=\sum_{i=1}^N \frac{\bp_i^2}{2}+\frac12 \sum_{i=1}^N \bx_i^2$ 
satisfying the boundary conditions corresponding to the two fixed points. 

The energies of these states are 
\begin{equation}
 \Delta_N^{(+)}=N+\frac{1}{2k}N(N-1)\,,
 \label{eq:chiral-plus}
\end{equation}
and
\begin{equation}
 \Delta_N^{(-)}=N-\frac{1}{2k}N(N-1)\,.
 \label{eq:chiral-minus}
\end{equation}
The ``chiral'' state
$\Psi_{N,+}$ coincides with the ground state at $k=\infty$, and as will be discussed, it remains the ground state in some range of parameters.  

There are many other exactly known eigenstates with higher energies than those of the states in~\eqref{eq:chiral-anyon-gauge}. These are usually called
\emph{linear states} because their energies depend linearly on $\frac{1}{k}$~\cite{Khare,Doroud2016,Sen1992}. (Note that for two particles all states are linear~\eqref{exacttwobody}.) We will not discuss these states further, for now. 
Just note that the wave function $\Psi_{N,-}$ with $\frac{N}{k}>2$ is non-normalizable because, when all relative coordinates are scaled to zero, its norm
contains
\begin{equation}
 \int \dd s\;s^{2(N-1)-1-\frac{1}{k} N(N-1)}\,.
\label{eq:singular-hyperradial-norm}
\end{equation}
This integral diverges if
$\frac{N}{k}>2$. %
Therefore, the naive dimensions of the operators 
$(\Phi^\dagger)^N$ drop below the unitarity bound for any finite $k$ at large enough $N$ at the unstable ``fixed point''~\cite{Doroud2018}. What happens for $\frac{N}{k}>2$ is that new linear states become normalizable. 
Immediately above $\frac{N}{k}=2$, the normalizable linear state with the lowest energy is
\begin{equation}
\widetilde \Psi_{N,-}^{(-2)}\propto r^{\frac{N(N-1)}{k}-2N}\sum_{i<j}(\bar z_i-\bar z_j)^2\,\Psi_{N,-}\,,\qquad r^2=\frac1N\sum_{i<j}|z_i-z_j|^2\,.
\end{equation}
Its energy
\begin{equation}
\widetilde\Delta_N^{(-)}
 =2-N+\frac{N(N-1)}{k} 
\end{equation}
is linearly increasing and coincides with $\Delta_N^{(-)}$ at $\frac{N}{k}=2$, so that the operator dimensions remain above the unitarity bound. We refer the reader to Appendix~\ref{3anyonsapp} for details. In that appendix, we also argue for the existence of an infinite set of unstable fixed points that differ by their multi-anyon contact interactions.

\subsection{Angular Momentum} 

Since the harmonic trap preserves rotational symmetry, we can discuss the angular momentum. 
We denote the physical total angular
momentum by $J$. There are essentially two contributions to the angular momentum; one comes from the orbital motion of anyons and the other from the statistics. %

We can always write the anyon wave function as  \begin{equation}\label{magneticgauge}
 \Psi_{\rm A}=\mathcal U\Psi_{\rm B}\,,
 \qquad
 \mathcal U=
 \exp\!\left[\frac{\ii}{k}\sum_{a<b}\arg(z_a-z_b)\right]\,,
\end{equation}
with $\Psi_{\rm B}(z_1,\ldots,z_N)$ a single-valued wave function. This is often called ``the magnetic gauge'', whereas $\Psi_{\rm A}$ is in the ``anyonic gauge''. The anyonic gauge wave function is more physical and more general, but the magnetic gauge wave function is often convenient for calculations in the harmonic trap.  
Under a rotation $z_a\mapsto e^{\ii\vartheta}z_a$, 
\[
 \mathcal U(e^{\ii\vartheta}z_1,\ldots,e^{\ii\vartheta}z_N)
 =e^{\frac{\ii}{k}\vartheta N(N-1)/2}\mathcal U(z_1,\ldots,z_N)\,.
\]
This phase comes about because a $2\pi$ rotation induces 
$N(N-1)$ elementary
exchanges, each with the phase (\ref{exchangephase}). Because $\Psi_{\rm B}$ is single valued, an angular-momentum eigenstate obeys
$\Psi_{\rm B}(e^{\ii\vartheta}z_a)=e^{\ii j\vartheta}\Psi_{\rm B}(z_a)$ with orbital angular momentum $j\in\mathbb Z$. So far we have the angular momentum  
\[
j+\frac{N(N-1)}{2k}\,.
\]
But this is not the final answer. 

We have to remember that each anyon is a little ribbon with internal spin $1/2k$. This gives an additional $\frac{N}{2k}$ units of internal angular momentum. 
Altogether %
we finally obtain 
\begin{equation}\label{angmatexact} J=j+\frac{ N(N-1)}{2k}+\frac{ N}{2k}
=j+\frac{ N^2}{2k}\,,\qquad 
j\in\mathbb Z\,.\end{equation}
Essentially, the content of this assertion is that the fractional part of the total angular momentum is known and is independent of the state or dynamics. 

Another way to derive this result~\eqref{angmatexact} is as follows: from the TQFT's perspective, we are discussing a state with $N$ Wilson lines, each describing a $\Phi$ particle. Fusing them, we obtain a Wilson line of charge $N$. The spin of this line in the TQFT is $\frac{N^2}{2k}$. The fractional part of this quantity is robust.

The chiral state~\eqref{eq:chiral-anyon-gauge} has $j=0$.
A very special class of states we will encounter later consists of screened states. In screened states the orbital angular momentum almost cancels the angular momentum in the braiding statistics
$j\simeq- N^2/2k$.

As we vary $k$, various anyon eigenstates with different $j\in \mathbb{Z}$ cross. This is clearly seen in numerical calculations of the eigenstates for $N=3,4$ 
~\cite{Khare,Mashkevich1994,Sporre1992}.
For instance, for $N=3$,
the fermion ground state has $j=-3$, while the boson ground state has $j=0$ and they cross at approximately
 $\frac{1}{k}\simeq0.71$~\cite{Mashkevich1994}.
There is one ground-state crossing for $N=3$. We will see that for large $N$ there are many crossings.

For large $N$ the chiral wave function, which is the boson ground state~\eqref{eq:chiral-anyon-gauge}, does not even have parametrically correct energy.
Indeed, its energy scales as $N^2$ while the ground state at the fermion point has energy scaling as $N^{3/2}$. This is a quick way to see that some crossing must occur. 

A general bound on the energies of the system is given in~\cite{Lundholm2017}. Denote by
$E_{0,j}$ the lowest energy in a fixed sector $j\in\mathbb{Z}$. Then
\begin{equation}
 E_{0,j}\geq
 N+\left|j+\frac{N(N-1)}{2k}\right| .
 \label{eq:trap-bound}
\end{equation}
To find the actual ground state at some fixed $k$ we then have to minimize $E_{0,j}$ over $j$.

We know that in the large-$N$ limit, in generic local interacting nonrelativistic systems, $\min_j\{E_{0,j}\}\sim N^{3/2}$.
Then, from the bound~\eqref{eq:trap-bound} it follows that the number of level crossings must grow with $N$~\cite{ChitraSen1992,Lundholm2017}.
In Appendix~\ref{sec:crossing-density} we derive an explicit formula for the density of crossings in our double-scaling limit.

\section{Double Scaling and the Solid Phase}
\label{sec:double-scaling}

As stated in the introduction, the system of anyons admits the limit \begin{equation}
 k\to\infty\,,
 \qquad N\to\infty\,,
 \qquad \lambda=\frac Nk\quad\text{fixed}\,,
 \label{eq:double-scaling}
\end{equation}
which renders the system semiclassical for all values of $\lambda$. 
This is achieved by rescaling the Chern--Simons matter action: \begin{equation}
 S=\int\dd t\,\dd^2x\left[
 \ii\Phi^\dagger D_t\Phi
 -\frac12(D_i\Phi)^\dagger D_i\Phi
 -\frac{ r^2}{2}\Phi^\dagger\Phi
 -\frac{\pi}{k}\Phi^\dagger\Phi^\dagger\Phi\Phi
 +\frac{k}{4\pi}\epsilon^{\mu\nu\rho}
 A_\mu\partial_\nu A_\rho
 \right]\,.
 \label{eq:CS-action}
\end{equation} To make the semiclassical limit manifest, we take
$\Phi=\sqrt{k}\,\phi$ with $\phi=\sqrt\rho\,e^{-\ii\chi}$.
The action and energy are proportional to $k$, see (\ref{eq:intro-density-action}), so saddle points are under control (quantum corrections due to the change to polar variables are negligible in this limit). 
We obtain the Hamiltonian  
\begin{equation}
 \frac{H}{k}=\int\dd^2x\left[
 \frac{(\nabla\rho)^2}{8\rho}
 +\frac{\rho}{2}(\nabla\chi-\bA)^2
 +\pi\rho^2+\frac{ r^2}{2}\rho\right]\,,
 \qquad
 \int\dd^2x\,\rho=\lambda\,.
 \label{eq:classical-energy}
\end{equation}
We also have the Gauss law constraint
$B=\epsilon^{ij}\partial_iA_j=2\pi\rho$.
The term $(\nabla \rho)^2 / \rho$ is sometimes called quantum pressure, although it is an essential part of the classical limit.
The Lagrangian in these variables is:
\begin{equation}
 L=k\int\dd^2x\left[
 \rho(\dot\chi-A_0)
 -\frac{1}{4\pi}\epsilon^{ij}A_i\dot A_j
 +\frac{A_0}{2\pi}B-\frac{\mathcal H}{k}
 \right]\,.
 \label{eq:full-time-action}
\end{equation}

The angular momentum for classical solutions can be calculated as follows:
\begin{equation*}\frac{1}{k}J_{\rm cl}=-\int\dd^2x\,\rho\,\left(\partial_\varphi\chi+A_\varphi\right).\end{equation*}
In Coulomb gauge this becomes 
\begin{equation}
   \frac{1}{k}J_{\rm cl}
 =-\int\dd^2x\,\rho\,\partial_\varphi\chi
 +\frac{\lambda^2}{2} = %
\lambda\left(W(\lambda)+\frac{\lambda}{2}\right)\,.
\label{eq:double-scaling-angular-momentum}
\end{equation}
(We defined the function $W(\lambda)$ for future reference.) These classical expressions do not respect
$j\in\mathbb{Z}$ even though the term 
$N^2/2k$ does come out exactly, as in~\eqref{angmatexact}. To ensure $j\in\mathbb{Z}$ we need to quantize the rotor mode around the classical solution, as usual (see Appendix~\ref{sec:crossing-density}).

\subsection{Vortex-Free Solutions and the Taubes Equation}

An interesting class of solutions consists of those with $\partial_\varphi\chi=0$: no vortices are present and $\rho$ has a smooth density profile. 
The chiral wave function~\eqref{eq:chiral-anyon-gauge} is an example of such a state. As a warm-up, it is useful to see how it arises in the double-scaling limit from the saddle point equations of~\eqref{eq:classical-energy}. 
Assuming no vorticity and time independence, we can derive the equation of motion for $\rho$.
In Coulomb gauge \(\nabla\cdot\bA=0\) the Gauss law is solved by
\begin{equation}
	A_i= -\epsilon_{ij}\int\dd^2y\,\frac{(\bx-\bm y)_j}{|\bx-\bm y|^2}\rho(\bm y).
\end{equation}
Therefore, assuming no vorticity, we minimize
\begin{equation}
	\frac{E}{k} = \int\dd^2x\left[\frac{(\nabla\rho)^2}{8\rho} +\frac{\rho}{2}A_i^2 +\pi\rho^2+\frac{ r^2}{2}\rho \right].
\end{equation}
at fixed
\begin{equation}
	\lambda=\int\dd^2x\,\rho\,.
\end{equation}
A trick to minimize the energy is to rewrite the energy functional as\footnote{To see that this holds, expand the last term as follows:

\begin{align}\label{nov}
\frac{1}{2}&
\int\dd^2x\,\rho
\left(
\frac{1}{2}\partial_i\log\rho
-\epsilon_{ij}A_j+x_i
\right)
\left(
\frac{1}{2}\partial_i\log\rho
-\epsilon_{ik}A_k+x_i
\right)\\
=&
\int\dd^2x
\left[
\frac{(\nabla\rho)^2}{8\rho}
+\frac{\rho}{2}A_iA_i
+\frac{r^2}{2}\rho
\right]
-\frac{1}{2}
\int\dd^2x\,
(\partial_i\rho)\epsilon_{ij}A_j
+\frac{1}{2}
\int\dd^2x\,x_i\partial_i\rho
-\int\dd^2x\,\rho x_i\epsilon_{ij}A_j\,.\nonumber
\end{align}
The first cross term is
\begin{align}
-\frac{1}{2}
\int\dd^2x\,
(\partial_i\rho)\epsilon_{ij}A_j
&=
\frac{1}{2}
\int\dd^2x\,
\rho\epsilon_{ij}\partial_i A_j
=\pi\int\dd^2x\,\rho^2\,.
\end{align}
The second cross term is
\begin{equation}
\frac{1}{2}
\int\dd^2x\,x_i\partial_i\rho
=
-\int\dd^2x\,\rho
=
-\lambda\,.
\end{equation}
The final cross term, when written in terms of the density, reduces to 
\[
\int\dd^2x\,\rho(\bx)
x_i\epsilon_{ij}A_j(\bx)=\frac12\int\dd^2x\,\dd^2y\,
\rho(\bx)\rho(\bm y)
=
\frac12\lambda^2\,.
\]
} 
\begin{equation}\label{TaubesFun}
\frac{E[\rho]}{k}
=
\lambda+\frac{\lambda^2}{2}
+
\frac{1}{2}
\int\dd^2x\,\rho
\left(
\frac{1}{2}\partial_i\log\rho
-\epsilon_{ij}A_j+x_i
\right)^2~.
\end{equation}

The minimum satisfies 
$\frac{1}{2}\partial_i\log\rho
-\epsilon_{ij}A_j+x_i=0$.
Taking the divergence and using the Gauss law gives
\begin{equation}
\nabla^2\log\rho=4\pi\rho-4\,.\label{Taubes}
\end{equation}
In terms of $h=\log\bigl(\pi\rho\bigr)$,
the density equation becomes
\begin{equation}
	\nabla^2 h=4\left(e^h-1\right).
\end{equation}
This is the source-free Taubes equation. 

Let us see how the anyon wave function~\eqref{eq:chiral-anyon-gauge} obeys this equation. To infer the density we consider 
\begin{equation}
\bigl|\Psi_{N,+}\bigr|^2\propto
 e^{-\sum_i|z_i|^2}\prod_{i<j}|z_i-z_j|^{2/k}\,.
 \label{chiraldensity}
\end{equation}
We need to determine the density $\rho(\bx)=\frac{1}{k}\left\langle
\sum_{i=1}^{N}\delta^{(2)}(\bx-\bx_i)
\right\rangle$ that follows from this wave function. The density is normalized with an additional factor of $1/k$ as in the rest of the text. 
We can interpret the modulus of the wave function~\eqref{chiraldensity} as the Boltzmann weight of a two-dimensional
Coulomb gas. The corresponding {\it auxiliary} continuum energy functional for this Boltzmann gas is
\begin{equation}
\frac{\mathcal E}{k}
=
\int\dd^2x\,\rho(\bx)|\bx|^2
-
\int\dd^2x\,\dd^2y\,
\rho(\bx)\rho(\bm y)\log|\bx-\bm y|+\int\dd^2x\,\rho(\bx)\log\rho(\bx)\,.
\end{equation}
Here, the last term is an entropy factor originating from the change of variables \cite{Dyson:1962es}.\footnote{There is an additional term that comes from $\sum_{i\ne j}\log|z_i-z_j|=\sum_{i,j}\log|z_i-z_j|+\frac12\sum_i\log\rho(\bx_i)$~\cite{Wiegmann:2005eh}. However, it is proportional to $1/k$ and therefore is subleading in our limit. The entropy factor is derived as follows: we divide space into small cells with
occupation numbers $n_a$, where $\sum_a n_a=N$. The number of
microscopic configurations corresponding to these occupation numbers is
\begin{equation}
\frac{N!}{\prod_a n_a!}\,.
\end{equation}
Using Stirling's approximation and taking the continuum limit, one finds that we need to add the Jacobian (using our convention for the normalization of the density)
\begin{equation}
\exp\left[
-k\int\dd^2x\,\rho(\bx)\log\rho(\bx)
\right]\,,
\end{equation}
up to density-independent factors.}
Minimizing this functional subject to the constraint
$\int\dd^2x\,\rho(\bx)=\lambda$ gives
\begin{equation}
|\bx|^2
-2
\int\dd^2y\,\rho(\bm y)\log|\bx-\bm y|
+
\mu_{\rm L}+\log \rho
=
0\,,
\end{equation}
where $\mu_{\rm L}$ is a Lagrange multiplier. Applying the two-dimensional
Laplacian,
we obtain
precisely equation~\eqref{Taubes}. Therefore the chiral wave function is indeed compatible with our semiclassical limit.

Let us seek solutions $\rho(r)$ that are functions only of the distance. 
At large $r$ we have $\rho(r)\sim r^{2\lambda}e^{-r^2}$.

For $\lambda\ll1$, the density--density interaction is subleading and we just have to solve a Poisson-like equation. To leading order, the normalized density is Gaussian:
\begin{equation}
\rho(r)=\frac{\lambda}{\pi}e^{-r^2}+\ldots\,.
\end{equation}
For $\lambda\gg1$, 
the problem simplifies since the entropy term goes away and we simply obtain a constant density up to the radius $R=\sqrt\lambda$:
\begin{equation}
\label{largelamtau}
\rho(r)=\frac1\pi\Theta(\sqrt\lambda-r)+\ldots\,.
\end{equation}
This constant density trivially solves the Taubes equation~\eqref{Taubes} away from the boundary. The energy is $\frac{E}{k}=\lambda+\frac{\lambda^2}{2}$, which agrees with the exact energy~\eqref{eq:chiral-plus} $\frac{E}{k}=\lambda+\frac{\lambda^2}{2}-\frac{\lambda}{2k}$ in the double-scaling limit.
The crossover between the two types of profiles is similar to the analysis of the Coulomb gas in~\cite{AkemannByun2019}.

As explained in the previous section, the solution of~\eqref{Taubes} is not the ground state in the thermodynamic limit. 
Next, we will see that within the double-scaling limit it is the ground state only up to some finite value of $\lambda\sim 2.96$, beyond which it is replaced by solutions with vorticity. Among solutions with no vorticity (\emph{i.e.}, a constant phase $\chi$), the energy cannot be lowered, as is manifest in~\eqref{nov}. In particular, the large-$\lambda$ solution of the Taubes equation~\eqref{largelamtau}, where the profile is nearly constant in a very large droplet, is far from the ground state and essentially irrelevant for the low-energy physics.

\subsection{Axisymmetric Solutions with Vorticity}\label{sec:Axisymmetric_saddles}

The simplest saddle with vorticity is a vortex of integer winding at the center of the trap. Here we study these rotationally invariant solutions. They provide an explanation for the first level crossing, where the chiral solution of~\eqref{Taubes} ceases to be the ground state. This level crossing is due to the formation of one elementary vortex at the center of the trap. 

For an axisymmetric saddle, we take
\begin{equation}
\chi=\ell\varphi\,,\qquad\bA=A_\varphi(r)\,\widehat\varphi\,,\qquad
 \ell\in\mathbb Z_{\geq0}\,.\label{eq:radial-ansatz}
\end{equation}
Regularity at the origin requires $\rho(r)\propto r^{2|\ell|}$. Indeed, a vortex produces a zero of the density. 
Writing $q=\sqrt\rho$, the energy~\eqref{eq:classical-energy} reduces to the
one-dimensional functional
\begin{equation}
 \frac{E}{2\pi k}
=\int\limits_0^\infty\dd r\,r\left[
 \frac12(q')^2
 +\frac{q^2}{2r^2}(\ell-Q)^2
 +\pi q^4
 +\frac{r^2}{2}q^2
 \right]\,,
 \qquad
 \frac{\dd Q}{\dd r}=2\pi r q^2,
 \qquad Q(\infty)=\lambda\,.
\label{eq:radial-functional}
\end{equation}
This functional is slightly nonlocal due to the appearance of the function $Q(r)$, which measures the total number of particles inside the disk of radius $r$. 

The solution in Eq.~\eqref{largelamtau} is certainly not the ground state at large $\lambda$. 
To see this, we simply plug the ansatz~\eqref{largelamtau} into the energy functional~\eqref{eq:radial-functional}
and obtain from the cross term a contribution proportional to
$\int\frac{\dd r}{r} q^2\ell Q\sim \ell\int r\dd r \Theta(\sqrt\lambda-r) \sim \ell \lambda$. It is therefore very clear that the energy can be lowered by allowing vortices (with the right sign of $\ell$). 

We minimize~\eqref{eq:radial-functional} at fixed $\lambda$. This must be done numerically. We find that as $\lambda$ increases, the dominant $\ell$ also increases. The first few level crossings among the axisymmetric solutions are reported in Table~\ref{tab:radial-envelope}.

\begin{table}[!ht]
\centering
\small
\begin{tabular}{cc}
\toprule
$\lambda$ interval & favored radial winding $\ell$ \\
\midrule
$(0,2.96)$ & 0 \\
$(2.96,6.17)$ & 1 \\
$(6.17,9.11)$ & 2 \\
$(9.11,11.93)$ & 3 \\
$(11.93,14.67)$ & 4 \\
$(14.67,17.36)$ & 5 \\
$(17.36,20.01)$ & 6 \\
\bottomrule
\end{tabular}
\caption{Level crossings among the axisymmetric solutions}
\label{tab:radial-envelope}
\end{table}

These axisymmetric solutions cannot be the ground states either! Physically, the vortices would like to split. A direct way to see the issue with these axisymmetric solutions is to calculate their energy as a function of $\lambda$. 
Let $\lambda_\ell$ denote the value of
$\lambda$ at which the preferred winding changes from $\ell$ to
$\ell+1$. We show in Appendix~\ref{Giant_vortex_app} that
\begin{equation}
 \lambda_\ell
 =
 \frac94\,\ell+o(\ell)
 ,\qquad
 \ell\longrightarrow\infty.
 \label{eq:large-ell-crossing}
\end{equation}
The energy is 
\begin{equation}
\frac{E_{\rm min}}{k}
 =
\frac5{18}\lambda^2+o(\lambda^2)
 ,\qquad  
 \lambda\longrightarrow\infty.
 \label{eq:large-lambda-radial-envelope-energy}
\end{equation}

Comparing this energy with~\eqref{TaubesFun} for the chiral state with vanishing winding, where we have $\frac{E}{k}=\frac12\lambda^2+\cdots$, we see that a central giant vortex is asymptotically favorable (over a state with vanishing vorticity) at large $\lambda$, but the energy still scales as $\lambda^2$ and does not agree with the expected $\lambda^{3/2}$ scaling of generic interacting nonrelativistic systems.

Indeed, we will see momentarily that the $\ell=0$ and $\ell=1$ axisymmetric states are true ground states in a range of $\lambda$, but the $\ell\geq 2$ vortices are all unstable to splitting. 

For the discussion below it is useful to elucidate the physical interpretation of these vortices. First of all, their angular momentum is $j=-\ell N$. From the decay of the density as $r^{2|\ell|}$, and comparing to the decay rate when two anyons with statistical angle $\theta$ approach each other, $r^{2\theta/\pi}$, it is clear that we should think of a vortex as  a ``collection'' of $k\ell$ anyons, with total magnetic flux $\ell$. This is also evident from the phase $\chi$ having circulation $2\pi\ell$. Thus, solutions with nonzero $\ell$ should be viewed as the usual Chern--Simons--Higgs vortices, which are qualitatively  made of $k$ elementary anyons. Outside the vortex, we are in an ordinary gapped Higgs phase of the Chern--Simons matter theory while the vortex is a massive excitation in this theory. This perspective will be very useful later.

\subsection{Splitting into Elementary Vortices}\label{splittingsec}
Above we have argued that giant vortices at the center of the trap cannot be the ground state, as they do not exhibit the thermodynamic behavior expected of $N$ interacting nonrelativistic particles.

Indeed, the $\ell\geq 2$ vortices of the previous section all possess  instabilities that are visible in the linearized approximation.  To see this, we expand the saddle points corresponding to general $\ell$ in small fluctuations. We organize the fluctuations into modes of well-defined angular momentum,
\begin{equation}q(r)e^{\ii\ell\varphi}+
\sum_m a_m(r)e^{\ii(\ell+m)\varphi}
 +b_m(r)e^{\ii(\ell-m)\varphi}\,.
 \label{eq:hessian-perturbation}
\end{equation}
Since the original saddle for the central $\ell$ vortex is known only numerically, the fluctuation analysis is also numerical, with the following sample of results:
\begin{table}[!ht]
\centering
\small
\begin{tabular}{ccccc}
\toprule
$\ell$ & $\lambda$ & $m$ & lowest eigenvalue & conclusion \\
\midrule
$0$  & $1.48$  & $1$  & $+1.131$   & stable \\
$1$  & $4.56$  & $1$  & $+0.217$   & stable \\
$2$  & $7.64$  & $2$  & $-4.668$   & splits \\
$3$  & $10.52$ & $3$  & $-10.743$  & splits \\
$5$  & $16.01$ & $5$  & $-20.124$  & splits \\
$10$ & $29.08$ & $10$ & $-37.943$  & splits \\
$20$ & $54.37$ & $20$ & $-68.361$ & splits \\
\bottomrule
\end{tabular}
\caption{Some eigenvalues of selected modes.}
\label{tab:hessian-spectrum}
\end{table}

Thus the calculation shows that the $\ell=0,1$ saddles at small enough $\lambda$ are stable, while for all $\ell\geq 2$ there is an instability. 
For instance, two elementary vortices away from the origin dominate the single vortex starting at $\lambda\sim 5.1$ and the double vortex at the center with $\ell=2$ is never the ground state. 

We still have to find the correct minimum as $\lambda$ increases. Initially, as we remarked above, it is a state made of two elementary $\ell=1$ vortices. As we increase $\lambda$, the number of elementary vortices is expected to be large and a locally periodic lattice becomes the natural variational ground state. The harmonic trap breaks translational symmetry; thus, it is more precise to say that the anyons form a periodic lattice in a large hard trap, while in the harmonic trap the lattice is slightly deformed. 

\subsection{The Vortex Crystal}
\label{sec:vortex-crystal}

We now derive the large-$\lambda$ energy and density profile. A key question is how many ``free anyons'' we have compared with ``anyons that are lumped  with vortices'' 
\begin{equation}
 \eta\equiv\frac{\rho_v}{\bar\rho}\,,
 \label{eq:cell-scales}
\end{equation}
where $\bar\rho$ is the
local mean total matter density (normalized by $1/k$ as mentioned already a few times) and $\rho_v$ is the number of unit vortices per
unit area.

We will see that in the $\lambda\to\infty$ limit all free anyons disappear and $\eta\to1$. Since the vortices are mutually local the system can be said to be screened in the sense that the statistical interactions vanish because each vortex carries a properly quantized magnetic flux.

The parameter $\eta$ controls the vorticity of the velocity $\bv$. More precisely, let
\begin{equation}
 \bv=\nabla\chi-\bA.
\end{equation}
The phase $\chi$ has circulation $2\pi$ around every unit vortex, while the Gauss law, after coarse graining,
gives $\nabla\times\bA=2\pi\bar\rho$. Therefore,
\begin{equation}
\nabla\times\bv=2\pi(\rho_v-\bar\rho)\,.
\end{equation}
Thus $\eta=1$ is precisely the condition that the statistical magnetic field
is screened by vortex circulation on average. 

The problem of finding the true ground state at large $\lambda$ is complicated by the fact that the quantum pressure term $\sim \frac{(\nabla\rho)^2}{\rho}
$ is not negligible. To see this, we drop the
quantum pressure and impose $\bv=0$ (anticipating screening) to obtain the following energy to minimize:
\begin{equation}
\int\dd^2x\left(\pi\rho^2+\frac{r^2}{2}\rho\right).
 \label{eq:smooth-functional}
\end{equation}
This is essentially a Thomas--Fermi approximation. Minimization at fixed $\int\dd^2x\,\rho=\lambda$ yields
\begin{equation}
 \rho(r)=\frac1{2\pi}
 \left(\sqrt{2\lambda}-\frac{r^2}{2}\right)\,,
 \qquad
 \frac{E}{k}=\frac{\sqrt8}{3}\lambda^{3/2}\,,
 \label{eq:smooth-result}
\end{equation}
for $r<(8\lambda)^{1/4}$. This is already a triumph compared to~\eqref{eq:large-lambda-radial-envelope-energy}, which did not even behave parametrically correctly. It certainly vindicates the screening ansatz. 
 
However, this result is not trustworthy since the omitted quantum pressure term is large. 
Indeed, the complete screening ansatz implies that the density of zeros is comparable to $\rho$. Therefore the distance between zeros should be $\sim 1/\sqrt\rho$.
Consequently $\nabla\rho\sim \rho^{3/2}$ and hence the quantum pressure term scales as $\rho^2$, 
\emph{i.e.}, exactly like the terms retained in
Eq.~\eqref{eq:smooth-functional}. Therefore, it is inconsistent to assume complete screening and drop the quantum pressure term. 

This is intuitively obvious; complete screening implies vortices and hence a dense network of zeros of the density, leading to large gradients.

\paragraph{The Unit Cell.}\label{sec:unit_cell}
To treat the quantum pressure correctly, consider a periodic lattice with mean matter
density $\bar\rho=\frac{1}{\eta}\rho_v$. We define the unit cell such that it contains one fundamental $\ell=1$ vortex. We do not assume complete screening; we will later show  that this occurs automatically at large $\lambda$. 

The area of the unit cell is $A=\rho_v^{-1}$. 
Minimizing the
microscopic fields within the unit cell defines an interesting quantity
\begin{equation}
 \widehat e(\eta)
 =\frac{1}{A\bar\rho^2}
 \min_{\rm cell}\int_{\rm cell}\dd^2x\left[
 \frac{(\nabla\rho)^2}{8\rho}
 +\frac{\rho}{2}\bv^2
 +\pi\rho^2
 \right].
 \label{eq:cell-energy-definition}
\end{equation}
The unit cell problem is solved at fixed $A^{-1}\int_{\rm cell}\rho=\bar\rho$, fixed $\rho_v=A^{-1}$, and a fixed $2\pi$ phase winding. (The Gauss law is also imposed in the unit cell.) 

On scales large compared with a unit cell, the energy becomes
\begin{equation}
 \frac{E_{\rm meso}}{k}
 =\int\dd^2x\left[
\frac12\bar \rho(\bar \bv^2+r^2)+\bar\rho^2\widehat e(\eta)
 \right]\,,\label{eq:mesoscopic-functional}
\end{equation}
where $\bar\bv$ is the average velocity over the unit cell.
In this sense, $\frac{E_{\rm meso}}{k}$
describes the smooth envelope only. Similarly, $\bar \bv$ describes some residual macroscopic vorticity. The
rapid density variation within each cell is captured in
$\widehat e$. This is the motivation for introducing $\widehat e$ in the first place.

It may seem tempting to conclude that
$\rho_v=\bar \rho$ and
$ \bar\bv=0$. This is the statement of complete screening, which holds only at infinite $\lambda$.
In Appendix~\ref{app:finite-lambda-screening} we find that, for instance, $\eta=0.95\pm0.03$ for
$\lambda=50$--$100$ (ignoring the edge of the system). Some matter remains that is not in the form of vortices near the edge, but in the bulk of the system, for $\lambda=50$--$100$, we already have essentially complete screening.

Therefore, to obtain the phase of the system at $\lambda\to\infty$, we set $\eta=1$. We find numerically that\footnote{Our value of $\widehat e(1)$ is consistent with recent numerical results in~\cite{AtaeiEtAl2025} at the infrared stable fixed point.}
\begin{equation}
\widehat e(1)=7.223\pm0.004\,.
\label{eq:cell-energy}
\end{equation}
The minimization strongly prefers equilateral triangles as unit cells. We will return to this point later.

\begin{figure}[t]
\centering
\includegraphics[width=0.85\textwidth]{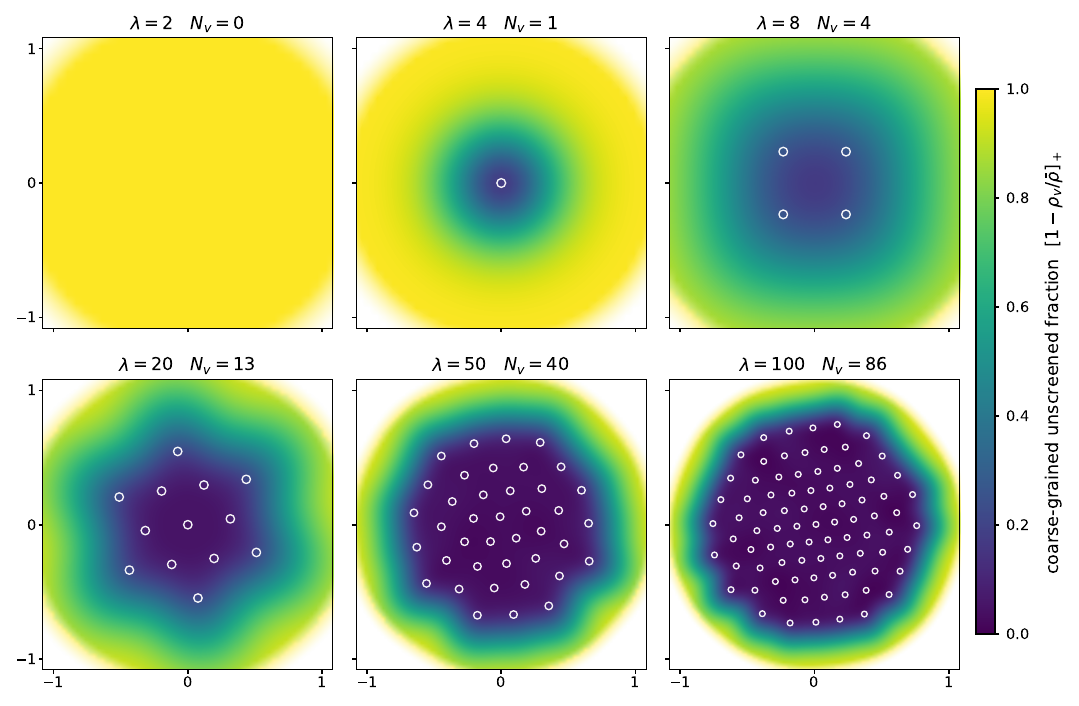}
\caption{Screening in the harmonic trap.  The yellow color is the
coarse-grained unscreened fraction
$1-\rho_{v}/\bar\rho$.  White rings mark the vortices.
As $\lambda$ grows, the unscreened component is depleted in the bulk
and survives mainly near the edge. We can see that the number of vortices approaches $\lambda$, albeit not very quickly, mostly due to edge effects.}
\label{fig:harmonic-screening-sequence}
\end{figure}

We schematically demonstrate how $\eta\to1$ is achieved at $\lambda\to\infty$ in Figure~\ref{fig:harmonic-screening-sequence}. We see that as we increase $\lambda$, we obtain a triangular lattice with essentially no matter outside of the vortices.

Having calculated~\eqref{eq:cell-energy}, we can now readily use~\eqref{eq:mesoscopic-functional} to obtain the coarse-grained density\footnote{Note that in the naive approach where the quantum pressure is dropped, the coefficient of $\rho^2$ is $\pi$; see~\eqref{eq:smooth-functional}. The quantum pressure term effectively replaces that coefficient by $\widehat e(1)\sim 7.22$.}

\begin{align}
 \bar \rho(r)
 &=\frac1{2\widehat e(1)}
 \left[\sqrt{\frac{2\widehat e(1)\lambda}{\pi}}
       -\frac{r^2}{2}\right]\Theta(R-r)\,,
 \label{eq:lattice-density}\\
 R
 &=\left(\frac{8\widehat e(1)\lambda}{\pi}\right)^{1/4}
 =2.07\,\lambda^{1/4},
 \label{eq:lattice-radius}\\
 \frac{E_0}{k}
 &=\frac23\sqrt{\frac{2\widehat e(1)}{\pi}}\,\lambda^{3/2}
 =1.43\lambda^{3/2}.
 \label{eq:lattice-energy}
\end{align}
(The coefficients above are determined to precision better than 1\%.)
As we mentioned earlier, the vortices have exchange statistics \((-1)^k\): bosonic for even \(k\) and fermionic for odd \(k\).
Since we are now dealing with a standard crystal state, the result is entirely insensitive to this distinction because localized lattice sites are basically frozen in a crystal.

We can also calculate the angular momentum of the solid at large $\lambda$.
Recall that the angular momentum must take the form 
\begin{equation}
 J=j+\frac{N^2}{2k},
 \qquad j\in\mathbb Z.
 \label{eq:angular-momentum-recalled}
\end{equation}
In a completely screened state, we can evaluate the first term of~\eqref{eq:double-scaling-angular-momentum} as \(W(\lambda) = - \lambda/2\). 
Therefore $J_{{\rm saddle}}/k$ vanishes to leading order. (This is unlike the giant-vortex states, which have $J_{{\rm saddle}}/k\sim \lambda^2$.) Another way to see that the angular momentum vanishes to leading order is to recall that  $\bar v=0$.
From this, we conclude that $J_{{\rm saddle}}/k$ is of order $1/k$.\footnote{We can also understand this cancellation microscopically.
Let
\begin{equation}
 \bA_i=\sum_{m\ne i}
 \frac{\widehat{\bm z}\times(\bx_i-\bx_m)}{|\bx_i-\bx_m|^2}
 \label{eq:microscopic-statistical-potential}
\end{equation}
be the statistical gauge field felt by particle $i$.
In the magnetic gauge, the wave function is 
$\Psi_{\rm B}=|\Psi_{\rm B}|e^{\ii S}$ with $\nabla_iS\simeq-\bA_i/k$, which follows from complete screening.

We can then calculate the angular momentum microscopically using 
\begin{equation}
 \left[\bx_i\times
 \frac{\widehat{\bm z}\times(\bx_i-\bx_m)}{|\bx_i-\bx_m|^2}
 +\bx_m\times
 \frac{\widehat{\bm z}\times(\bx_m-\bx_i)}{|\bx_i-\bx_m|^2}
 \right]_z=1.
 \label{eq:pair-angular-identity}
\end{equation}
Summing over the $N(N-1)/2$ pairs therefore gives that the leading term in the angular momentum cancels out since 
\begin{equation}
  \langle j\rangle_{\rm crystal}
 \simeq-\frac{N^2}{2k}+\cdots~,
\label{eq:crystal-angular-momentum}
\end{equation}
Therefore, $\langle J\rangle_{\rm crystal}$ vanishes to leading order, which means that it is at most of order $\mathcal O(1)$ in the double scaling limit (in the bulk of the crystal).}

The rotational zero mode has a simple consequence. At finite \(N\), the crystal orientation must be quantized. For a slow rigid rotation of a given classical saddle with angular velocity \(\Omega\), the angular momentum and energy are, to leading order,
\begin{equation}
  J=J_{\rm cl}+I\Omega+\cdots,
  \qquad
  E=E_{\rm cl}+\frac{I\Omega^2}{2}+\cdots ,
\end{equation}
where the moment of inertia \(I\) is of order \(N\). The corresponding quantum states form the usual Anderson tower. Using the relation between the total angular momentum and the integer angular momentum \(j\) of the wave function, their energies are
\begin{equation}
  E_j(\lambda)=E_{\rm cl}(\lambda)
  +\frac{\bigl(j-NW(\lambda)\bigr)^2}{2I(\lambda)}
  +\cdots\,.
\end{equation}
The classical saddle does not in general correspond to a state of definite angular momentum. Rather, it fixes the center of a tower of states labeled by \(j\in\mathbb Z\), and the ground state is obtained by choosing the integer closest to \(NW(\lambda)\).

As \(\lambda\) is varied, the center of the tower moves continuously. Two neighboring angular-momentum sectors cross whenever
\begin{equation}
  NW(\lambda_n)=n+\frac12 ,
  \qquad n\in\mathbb Z .
\end{equation}
This immediately gives an estimate for the frequency of the crossings 
\begin{equation}
  \frac{d n_{\rm cross}}{d\lambda}
  =
  N\left|W'(\lambda)\right| .
\end{equation}
These crossings are parametrically more numerous than the changes of the classical vortex configuration: over a finite interval of \(\lambda\), the former are of order \(N\), while the latter remain of order one.
In the fully screened regime the angular momentum found above gives
\begin{equation}
    \frac{d n_{\rm cross}}{d\lambda} \xrightarrow[\lambda \to \infty]{} \frac{N}{2} .
\end{equation}
More details are discussed in Appendix~\ref{sec:crossing-density}.

\subsection{Further Properties of the Crystal}
\label{sec:elasticity}
We now turn to the long-wavelength
mechanical response of the vortex crystal. On scales large compared with
the vortex spacing, but small compared with the size of the trap, the system behaves as an ordinary triangular crystal with mean matter
density $\bar\rho$. Complete screening implies that the vortex density is also
$\bar\rho$, so that the unit-cell area is $A_0=1/\bar\rho$.
Although the density follows a nonuniform Thomas--Fermi distribution, it varies slowly in the bulk, so its derivatives can be neglected (local density approximation).
In this approximation, the details of the confining potential enter only via the function \(\bar \rho\) and our results can be immediately extended to other configurations, including the hard trap of radius \(R\).

A slowly deformed crystal is described by a displacement field
$\bm u(\bx,t)$: a vortex whose equilibrium position is $\bx$ is moved to
$\bx+\bm u(\bx,t)$. The corresponding linear strain tensor and its
traceless part are
\begin{equation}
 u_{ij}=\frac12\left(\partial_i u_j+\partial_j u_i\right),
 \qquad
 u_{ij}^{\rm T}=u_{ij}-\frac12\delta_{ij}u_{kk}.
 \label{eq:strain-tensor}
\end{equation}
The trace $u_{kk}=\nabla\cdot\bm u$ describes a local dilation or
compression: to linear order, $\delta\rho=-\bar\rho\,u_{kk}$. The traceless
tensor $u_{ij}^{\rm T}$ describes a shear, which changes the shape of a unit
cell without changing its area. (The antisymmetric piece $\partial_{[i} u_{j]}$ rotates the lattice without stretching any bonds, so it cannot appear in the static elastic energy.) 
A two-dimensional symmetric strain tensor has three independent components: one compressional component, \(u_{kk}=u_{xx}+u_{yy}\), and two shear components, which may be taken to be \(u_{xx}-u_{yy}\) and \(2u_{xy}\). The sixfold rotational symmetry of the triangular lattice makes the two shear components degenerate at quadratic order and forbids their coupling to the compression. Consequently, although there are three independent strain components, the quadratic elastic energy is characterized by only two independent elastic constants: the bulk modulus \(K\) and the shear modulus \(G\). In this sense, the long-wavelength elasticity of the triangular lattice is isotropic at quadratic order. 
(The physical moduli are $k\bar\rho^2K$ and $k\bar\rho^2G$.) 

The bulk modulus $K$ is obtained directly from the energy stored in a unit cell~\eqref{eq:cell-energy-definition}.
By definition, the energy density of the unit cell is
\begin{equation}
 {\cal E}_{\rm cell}(\bar\rho)=k\widehat e(1)\bar\rho^2.
\end{equation}
Consequently, its pressure is
$P=\bar\rho\partial_{\bar\rho}{\cal E}_{\rm cell}-{\cal E}_{\rm cell}
=k\widehat e(1)\bar\rho^2$, and its physical bulk modulus is
$\bar\rho\partial_{\bar\rho}P=2k\widehat e(1)\bar\rho^2$. In our dimensionless
convention, this gives
\begin{equation}
 K=2\widehat e(1).
 \label{eq:bulk-modulus-definition}
\end{equation}

The shear modulus is obtained from a separate deformation of the periodic
unit-cell problem. If $\bm a_1,\bm a_2$ are the equilibrium lattice vectors,
we apply the area-preserving simple shear
\begin{equation}
 \bm a_\alpha\longmapsto F(\xi)\bm a_\alpha,
 \qquad
 F(\xi)=
 \begin{pmatrix}
 1&\xi\\
 0&1
 \end{pmatrix},
 \qquad
 \det F(\xi)=1.
 \label{eq:cell-shear}
\end{equation}
For each \textit{fixed} $\xi$, the fields $\rho,\chi,\bA$ are minimized again inside the
deformed periodic cell, at fixed mean density, with one unit of vortex
winding, and subject to the Gauss law. 
Then the shear modulus is defined through
\begin{equation}
\frac{E(\xi)}{kA_0\bar\rho^2}
 =\widehat e(1)+\frac12G\,\xi^2+O(\xi^4).
 \label{eq:shear-expansion}
\end{equation}

We can include $K$ and $G$ in the effective theory for the phonons: \begin{equation}
 H_{\rm el}^{(2)}
=\frac{k\bar\rho^2}{2}\int\dd^2x\left[
 K(u_{kk})^2+2G u_{ij}^{\rm T}u_{ij}^{\rm T}
 \right].
 \label{eq:elastic-hamiltonian-position}
\end{equation}
A numerical calculation of the unit cell energy with shear allows us to extract $G=1.354\pm0.004$. In summary, we find 
\begin{equation}
G=1.354\pm0.004\,,
 \qquad K=2\widehat e(1)=14.446\pm0.008\,.
\label{eq:elastic-moduli}
\end{equation}

These results determine the static restoring forces but not yet the
dynamics. To obtain the phonon frequencies, we must also determine the
kinetic term.
Due to complete screening, translations of the
periodic saddle behave exactly as in an ordinary crystal and there is no Berry term.%
\footnote{One can verify explicitly using the Gauss law that the contributions of the Berry term and the CS term cancel each other and no single-derivative-in-time terms remain.}
Thus, we expect a standard quadratic kinetic term that is completely fixed by the inertial mass density of the unit cell \(k \bar\rho\).
We therefore find two linearly dispersing phonons
\begin{equation}
\omega_L(q)=\sqrt{\bar\rho(K+G)}\,q+O(q^2)\,,
 \qquad
 \omega_T(q)=\sqrt{\bar\rho G}\,q+O(q^2)\,, \label{eq:linear-phonons}
\end{equation}
where $\omega_L$ and $\omega_T$ are inferred from~\eqref{eq:elastic-hamiltonian-position} when written in Fourier space.

For reference, at the center of the trap the phonon velocities are
\begin{equation}
v_L(0)\sim1.53\lambda^{1/4}\,,
 \qquad
 v_T(0)\sim 0.45\lambda^{1/4}\,.
\label{eq:central-sound-velocities}
\end{equation}
Note that these are the only light modes. There is no additional superfluid
Goldstone mode since a smooth phase rotation of the field $\Phi$ can be canceled by a flat connection for the dynamical statistical gauge field. This is why the large-$k$ state is an ordinary crystal rather than a supersolid. We emphasize that this crystalline state requires no Coulomb repulsion or other ordinary long-range force; it is stabilized entirely by the anyonic statistical interaction.

\section{Phase Diagram of the Anyon Gas}
\label{sec:melting}
We will now use the results discussed so far to gain insight into the phase diagram of the anyon gas.

The ordinary thermodynamic limit is taken at fixed $k$ and $N\to\infty$. To learn something about the thermodynamic limit from the limit we have solved, where $\lambda = N/k$ is fixed as $N\to\infty$, an extrapolation is necessary. 
This section presents the results of this extrapolation. 

As discussed in the introduction, around $k=1$ it is plausible (and numerically supported) that the anyon gas is a homogeneous superfluid. Conversely, our analysis suggests that at fixed $k\gg1$ and $N\to\infty$ we obtain an ordinary crystal. 
Therefore, we expect some $k_c$ where a first-order transition occurs between the solid and superfluid phases. 
We will use results from the previous section to estimate it. 

Our starting point is the zero-point motion in the crystal. When it becomes too large, the crystal is assumed to melt.
As in the previous section, let $\bm u$ be the displacement of a vortex-lattice site. For the triangular lattice, the nearest-neighbor spacing is fixed at equilibrium as
\begin{equation}
a_v^2=\frac{2}{\sqrt3\bar\rho}\,.
 \label{eq:lattice-spacing}
\end{equation}
We have two phonon modes; from~\eqref{eq:linear-phonons} we obtain the zero-point variance 
\begin{equation}
\left\langle|u_s(\bm q)|^2\right\rangle
 =\frac{1}{2k \bar\rho\omega_s(q)}\,, \label{eq:mode-zero-point-motion}
\end{equation}
where $s=T,L$ labels the transverse and longitudinal modes.

Since we only know the long-wavelength approximation to $\omega$, we use the Debye approximation: to estimate the average displacement of the sites, we integrate over the Brillouin zone. Crudely, we replace the Brillouin zone with a disk of width $q_D$ such that
$ \int_{|\bm q|<q_D}\frac{\dd^2q}{(2\pi)^2}=\bar\rho$, which gives
$q_D=2\sqrt{\pi\bar\rho}$.
Assuming~\eqref{eq:linear-phonons} continues to hold up to the cutoff $q_D$, we obtain
\begin{align}
 \left\langle|\bm u|^2\right\rangle
 &\simeq
 \sum_{s=L,T}\int\limits_{|\bm q|<q_D}
 \frac{\dd^2q}{(2\pi)^2}
 \frac{1}{2k \bar\rho\omega_s(q)}
 =\frac{1}{2\sqrt\pi\,k\bar\rho}
 \left(\frac1{\sqrt{K+G}}+\frac1{\sqrt G}\right).
 \label{eq:debye-displacement}
\end{align}
This calculation is crude because it is controlled by the ultraviolet region of the Brillouin zone, whereas the dispersion relation is known only in the infrared.\footnote{We have attempted, using GPT--5.6 Sol Codex, and Claude Code Fable 5.1 to perform the full integral over the Brillouin zone, including the nonlinearities of the dispersion, and obtained a numerical result very close (within 2\%) to our naive calculation.} 

An important heuristic parameter that controls the stability of solids is the Lindemann ratio, which measures the scale of fluctuations relative to the inter-site distance:
\begin{equation}
\gamma_{\rm L}^2
\equiv\frac{\langle|\bm u|^2\rangle}{a_v^2}
 \simeq
 \frac{\sqrt3}{4\sqrt\pi\,k}
 \left(\frac1{\sqrt{K+G}}+\frac1{\sqrt G}\right).
 \label{eq:lindemann-general}
\end{equation}
The result is independent of \(\bar \rho\), and remains valid for any choice of appropriate confining potential.

Plugging in the numerical values of $K$ and $G$ from the previous section~\eqref{eq:elastic-moduli}, we find \begin{equation}
\gamma_{\rm L}\simeq\frac{0.52}{\sqrt k}\,.
\label{eq:lindemann-number}
\end{equation}
The Lindemann ratio is the same throughout the bulk since it is a local property of the solid. Quantum Monte Carlo calculations for several two-dimensional triangular crystals give critical Lindemann ratios (on the crystalline side of the zero-temperature quantum melting transition) in the approximate range~\cite{Babadi2013,Astrakharchik2021}
$\gamma_{\rm L} \simeq 0.23\text{--}0.28$. This corresponds to\footnote{\label{anotherL}Perhaps one should not be satisfied with using $\langle|\bm u|^2\rangle$ in 2+1 dimensions since it diverges at finite temperature. This is because at finite temperature the calculation of $\langle|\bm u|^2\rangle$ is very similar to the VEV of the square of a free field in 2 Euclidean dimensions, which is known to diverge logarithmically. A quantity that avoids this difficulty is the relative displacement of two neighboring vortices~\cite{BruunNelson2014,BedanovGadiyakLozovik1985Lindemann,BedanovGadiyakLozovik1985Melting,Khrapak2020},
\begin{equation}
 \frac{\langle |\Delta\bm u|^2\rangle}{a_v^2}\,,
 \qquad
 \Delta\bm u=\bm u(\bm R+\bm a)-\bm u(\bm R)\,.
\end{equation}
The corresponding integral over the Brillouin zone is 
\begin{equation}
 \langle |\Delta\bm u|^2\rangle
 =\sum_{s=L,T}\int_{\rm BZ}\frac{\dd^2q}{(2\pi)^2}
 \frac{1-\cos(\bm q\cdot\bm a)}{k\bar\rho\omega_s(q)}\,.
 \label{eq:lindemann-notes-relative-bz}
\end{equation}
We can now define the relative Lindemann ratio
\begin{equation}
 \gamma_{\rm rel}^2
 \equiv\frac{\langle |\Delta\bm u|^2\rangle}{a_v^2}\,.
\end{equation}
Performing the integral over the Brillouin zone, we find  $\gamma_{\rm rel}\simeq0.621/\sqrt{k}$. Assuming melting occurs for $\gamma_{\rm rel}\simeq0.25\text{--}0.30$ gives an estimate very similar to the one previously obtained:
\begin{equation}
 k\sim4\text{--}6\,.
\end{equation}

} 
\begin{equation}
 k_c \simeq3.5\text{--}5.
 \label{eq:melting-estimate}
\end{equation}

There is another estimate, which is equally uncontrolled but leads to the same qualitative conclusion. 
In~\eqref{eq:lattice-energy} we obtained
\begin{equation}
 E_{\rm crystal}\simeq1.43\,\frac{N^{3/2}}{\sqrt k}\,.
 \label{eq:crystal-fixed-k-energy}
\end{equation}
This is derived in the double scaling limit, but again, we will use it away from the domain where it is exact.

Near the fermionic endpoint a complementary magnetic Thomas--Fermi theory is
available~\cite{Levitt2026}.  In terms of the effective statistical parameter of fermions, 
\begin{equation}
 k'^{-1}=1-k^{-1}\,,
\end{equation}
($k'\to\infty$ corresponds to ordinary fermions, while $k'=1$ corresponds to bosons)
the leading energy in a harmonic trap is 
\begin{equation}
 E_{{\rm superfluid}} =\frac{2\sqrt2}{3}\sqrt{c(k')}\,N^{3/2}+o(N^{3/2})\,,
 \qquad
 c(k')=1+k'^{-2}\bigl(1-\{k'\}\bigr)\{k'\}\,,
 \label{eq:magnetic-TF-energy}
\end{equation}
where braces denote the fractional part.  Equating the two leads to the estimate
\begin{equation}
k_c\simeq 2\text{--}3\,.\label{eq:leading-energy-match}
\end{equation}
This is in the same ballpark as~\eqref{eq:melting-estimate}, which bolsters our confidence that the underlying physical picture is correct.

Combining these two estimates and the one in footnote~\ref{anotherL}, we arrive at the prediction~\eqref{estintro}. 

\section{Conclusions}
\label{sec:conclusion}

We studied a finite-density gas of Abelian anyons at the repulsive, infrared-stable fixed point of nonrelativistic Chern--Simons matter theory.
Our analysis is controlled in the double-scaling limit
\begin{equation}
    k\to\infty,\qquad N\to\infty,\qquad
    \lambda\equiv \frac{N}{k}\ \text{fixed}~,
\end{equation}
in which \(1/k\) plays the role of an effective Planck constant.  
The many-body problem is thereby reduced to a semiclassical saddle-point problem
with a dynamical statistical gauge field.
  
At small \(\lambda\), the ground state is the vorticity-free  chiral droplet, and the full many-body wave function is known exactly in that regime.   
At \(\lambda\simeq2.96\), an elementary vortex appears at the center of the trap.  
As \(\lambda\) increases, further vortices appear in the droplet, while
centered vortices of winding \(\ell\geq2\) are unstable to splitting, so
at large \(\lambda\) the ground state is a lattice of elementary vortices.

The large-\(\lambda\) state is a triangular vortex crystal.  It exhibits complete screening: in the bulk, the density of vortices approaches the $1/k $ normalized coarse-grained matter density.  
The circulation of the vortices then cancels the statistical magnetic field on long distance scales.
The vortices are accompanied by \(k\) fundamental anyons. 
Since they carry an integer unit of statistical flux, their long-distance
statistical interactions are screened.  The resulting state is an ordinary
crystal.  In particular, its only light modes are the
longitudinal and transverse phonons. Let us stress again that this crystalline state requires no Coulomb repulsion or other ordinary long-range force; it is stabilized entirely by the anyonic statistical interaction.

We computed the energy of the unit cell and found that the triangular lattice is preferred.  
This leads to the large-\(\lambda\) energy
\begin{equation}
    \frac{E_0}{\omega}
    =1.43\frac{N^{3/2}}{\sqrt{k}},
\end{equation}
and we also extracted the elastic moduli of the crystal.  
We additionally discussed the angular-momentum structure of the semiclassical states.
At finite \(N\), each classical saddle gives rise to an Anderson tower, and the density of angular-momentum crossings is of order \(N\).

We then extrapolated our results to the thermodynamic limit, where $k$ is finite and $N$ is taken to infinity.
We propose a phase diagram that consists of a solid state for $k>k_c$ and a superfluid state for $k<k_c$.
We estimated $k_c$ using both an energy comparison and an analysis of the quantum fluctuations, and concluded that the transition is likely to occur around $k_c\simeq 2\text{--}5$.
We expect that the solid-fluid transition continues to hold at low nonzero temperature.

\medskip
There are several natural directions for future work.  
It would be interesting to study the system directly at fixed \(k\) using many-body
numerics.  We mentioned preliminary variational Monte Carlo results in footnote~\ref{prelimMonte}.  A systematic calculation at larger \(N\) could test the proposed phase diagram, determine the finite-size scaling of crystalline order, and decide whether the transition is a single first-order transition.

Furthermore, the melting estimate could be improved.  Our calculation uses the long-wavelength phonon spectrum together with a Debye approximation. However, the displacement fluctuations are sensitive to momenta near the Brillouin-zone scale.  It would be useful to compute the full phonon spectrum of the vortex crystal and use it to obtain a more reliable estimate of the quantum melting point.

The unstable fixed point with attractive anyons has a rich many-body structure that remains to be understood.  The attractive two-anyon boundary condition does not by itself determine the behavior when several anyons coincide.  
Appendix~\ref{3anyonsapp} finds, within the linear-state analysis, a web of possible fixed points associated with higher-body contact interactions, with operator dimensions remaining above the unitarity bound.  Whether these fixed points extend beyond the linear states and define complete unitary many-body theories is an open question.

Finally, it would be interesting to extend the analysis to finite temperature and other types of  anyons. The simplest generalization is when there are several species of Abelian anyons.

\paragraph{Acknowledgements.}
We thank Simeon Hellerman for collaboration at an early stage of this project. We thank Max~Metlitski, Senthil~Todadri, and Paul~Wiegmann for discussions. Z.K. and X.L. gratefully acknowledge NSF Award Number 2310283.
S.R.'s work was supported by the Swiss National Science Foundation under grant
number 200021\_219267. A.S. is supported by the Israel Science Foundation, grant number 1099/24. The authors used Codex by OpenAI and Claude Code by Anthropic to improve some figures, and to produce and double-check numerical code. %
\appendix

\section{Numerical Methods and Screening at Large $\lambda$}
\label{app:numerical_methods}
\subsection{Numerical Methods}
This appendix summarizes the numerical calculations and methodology used in the main text.

The axisymmetric giant-vortex saddles in Section~\ref{sec:Axisymmetric_saddles} are obtained by discretizing the radial energy functional~\eqref{eq:radial-functional} and minimizing it at fixed $\lambda$ using the L-BFGS-B algorithm~\cite{ByrdLuNocedalZhu}. We use a uniform radial grid on $[0,R_{\max}]$ with $N_r$ points. For the level-crossing calculation, we take $R_{\max}=12$ and $N_r=160,240,320,480$. The crossing values exhibit quadratic convergence in the grid spacing $a=R_{\max}/N_r$; we therefore fit $\lambda_{\ell}(a)=\lambda_{\ell}+c_{\ell}a^2$ and use the $a\to0$ intercepts as the continuum values quoted in the text. Finite-domain effects are checked at fixed $a=0.025$ using $(R_{\max},N_r)=(10,400),(12,480),(14,560)$, and the resulting values of $\lambda_\ell$ agree within $2\times10^{-9}$.

The vortex stability analysis in Section~\ref{splittingsec} is performed by perturbing the axisymmetric saddles in fixed angular sectors as in~\eqref{eq:hessian-perturbation}. The perturbed fields are normalized to the same $\lambda$ and the corresponding gauge fields are determined from the Gauss law. For each angular sector, we evaluate the  Hessian by centered finite differences (with step $\epsilon$) of the energy, using $R_{\max}=14$ and $(N_r,\epsilon)=(240,10^{-3}),(360,7\times10^{-4})$. For the sectors reported in Table~\ref{tab:hessian-spectrum}, the angular resolution is $N_\theta=64$ for $\ell\leq5$, $N_\theta=88$ for $\ell=10$, and $N_\theta=168$ for $\ell=20$. The radial functions $a_m(r)$ and $b_m(r)$ in~\eqref{eq:hessian-perturbation} are each expanded in a generalized-Laguerre basis. The number of radial basis functions is increased from $12$ in steps of $4$, up to $48$, until the lowest Hessian eigenvalue changes by less than $2\times10^{-3}$ upon a further basis enlargement.

The periodic unit-cell calculation of Section~\ref{sec:unit_cell} is performed in a rhombic cell with area $A=1$ and mean density $\bar{\rho}=1$, containing one unit vortex. The unit-cell energy functional~\eqref{eq:cell-energy-definition} is discretized on an $N\times N$ Fourier grid and minimized using the L-BFGS-B algorithm. The normalization of the density is fixed and the gauge field is determined from the Gauss law throughout the minimization. For the triangular cell, we use $N=15,25,41,65,97,129$. We also vary the rhombus opening angle $\alpha$ at $N=41$, using $\alpha=60^\circ,62^\circ,64^\circ,70^\circ,80^\circ,90^\circ$. We find that the minimum occurs at $60^\circ$, corresponding to the triangular lattice. To extrapolate the unit-cell energy to the continuum limit, we fit the results as functions of $1/N$ using several fitting ranges and polynomial orders. The resulting extrapolations give the quoted value $\widehat e(1)=7.223\pm0.004$,
with the uncertainty reflecting the spread among the fits. This determines the leading-order ground-state energy $E_0/k=(1.4296\pm0.0004)\lambda^{3/2}$ in the double-scaling limit; in the text we mostly quote the number 1.43 to be conservative.

The shear modulus calculation in Section~\ref{sec:elasticity} is performed by repeating the above unit-cell minimization after the area-preserving simple shear~\eqref{eq:cell-shear} of the triangular cell. We use $\xi=0.008,0.012,0.018,0.026$, including both signs, on grids $N=41,65,97,129$. Linear and quadratic continuum extrapolations in $1/N$ give $G=1.3532$ and $1.3550$, respectively. As a numerical check, we repeat the calculation using the area-preserving pure shear $F_{\rm p}(t)=\operatorname{diag}(e^t,e^{-t})$, with $t=0.008,0.012,0.018,0.026$ including both signs. This gives $G=1.3503$ and $1.3531$ for the linear and quadratic extrapolations, consistent with the simple-shear result. We quote in~\eqref{eq:elastic-moduli} $G=1.354\pm0.004$, using the simple-shear result as the central value. The uncertainty conservatively accounts for the continuum extrapolation and the variation between the two shear prescriptions.

\subsection{Towards Screening at Large \texorpdfstring{$\lambda$}{lambda}}\label{app:finite-lambda-screening}
We summarize in this subsection the numerical calculations that indicate complete screening at large $\lambda$.

We write the energy functional in terms of $\phi=\sqrt{\rho}\,e^{-\ii\chi}$ as
\begin{equation}
    \frac{E}{k}=\int\dd^2x\left[
    \frac12|(\nabla+\ii\bA)\phi|^2
    +\pi|\phi|^4+\frac{r^2}{2}|\phi|^2\right]\,,
    \qquad
    \int\dd^2x\,|\phi|^2=\lambda\,.
    \label{eq:finite-lambda-complex-functional}
\end{equation}
We discretize the complex field $\phi$ on a uniform $N\times N$ grid and minimize~\eqref{eq:finite-lambda-complex-functional} directly at fixed $\lambda$, using the L-BFGS-B algorithm.
The field $\phi$ is normalized at each step and the gauge field is reconstructed from the Gauss law $B=2\pi|\phi|^2$. Spatial derivatives are evaluated by Fourier differentiation. The initial seed contains either ring-like or approximately triangular arrangements of vortices. The vortex number is free to change. For the main minimizations, we take $N=97$ on $[-9,9]^2$. To check the grid dependence, we also use $N=129$ for $\lambda=20,50$, and $N=129,161$ on $[-10,10]^2$ for $\lambda=100$. Vortices are identified by the integer phase winding around elementary plaquettes. %

\begin{table}[ht]
\centering
\small
\begin{tabular}{@{}cccc@{}}
\toprule
$\lambda$ & $E/k$ & $N_v$ & $\eta=N_v/\lambda$\\
\midrule
$0.4$ & $0.47986$ & $0$ & $0$\\
$1$   & $1.4993$  & $0$ & $0$\\
$2$   & $3.9971$  & $0$ & $0$\\
$3$   & $7.4674$  & $1$ & $0.333$\\
$4$   & $11.256$  & $1$ & $0.250$\\
$6$   & $20.949$  & $2$ & $0.333$\\
$8$   & $32.155$  & $4$ & $0.500$\\
$10$  & $45.064$  & $5$ & $0.500$\\
$15$  & $82.740$  & $9$ & $0.600$\\
$20$  & $128.24$  & $13$ & $0.650$\\
$30$  & $234.24$  & $22$--$23$ & $0.733$--$0.767$\\
$50$  & $504.66$  & $40$--$42$ & $0.800$--$0.840$\\
$100$ & $1427.8$  & $85$--$86$ & $0.850$--$0.860$\\
\bottomrule
\end{tabular}
\caption{Energy and number of vortices in the lowest energy states for various values of $\lambda$.}
\label{tab:finite-lambda-screening}
\end{table}

We record the energies, total vortex numbers $N_v$, and screening ratios $\eta$ (estimated by $N_v/\lambda$) of the lowest energy states for various values of $\lambda$ in Table~\ref{tab:finite-lambda-screening}. Indeed, we find that for small $\lambda$ there are no vortices.  The centered one-vortex state follows.
By $\lambda\simeq15$ the zeros form the
beginning of a triangular lattice.
For the local screening diagnostic in Figure~\ref{fig:harmonic-screening-sequence}, we construct the vortex density $\rho_v$ from the vortex positions and coarse grain $\rho_v$ and $\rho$ with the same Gaussian kernel.
For configurations with several vortices, we use a Gaussian width $\sigma=0.72\,a_v$, with $a_v$ the
median nearest-neighbor vortex spacing. Figure~\ref{fig:harmonic-screening-sequence} shows $[1-\rho_v/\bar\rho]_+$ in the region where $\bar\rho$ exceeds $3.5\%$ of its maximum.
The values quoted above for $\eta$ underestimate the true screening effect because there are unscreened bosons near the edge of the droplet.\footnote{We do not go into details here, but the relative size of edge corrections is expected to scale as $\lambda^{-2/3}$.} For instance, if we ignore the edge and calculate the ratio of vortices to matter in the bulk, then we find that for $\lambda=50$--$100$, the actual bulk value of $\eta$ is $0.95\pm0.03$.

\section{More on the Unstable Fixed Point}
\label{3anyonsapp}

In Section~\ref{sec:contact}, we encountered serious difficulties with the unstable fixed point of the two-anyon interaction, $g=-\frac{2\pi}{k}$, which cast doubt on its existence as a genuine nonrelativistic fixed point. First, at fixed $k$, once the number of anyons exceeds a certain threshold, additional marginal and relevant operators appear and must be taken into account. Second, and closely related, when $N/k>2$, the wave function $\Psi_{N,-}$ in~\eqref{eq:chiral-anyon-gauge} ceases to be normalizable, and at the same time, the scaling dimension of the corresponding operator falls below the unitarity bound.

In this appendix, we revisit the unstable fixed point from the perspective of the linear states. We find evidence for a large web of fixed points and show that the dimensions of the operators associated with these states never violate the unitarity bound when the fixed points are analyzed properly.

We begin with the two-anyon sector, where the relevant structure can be exhibited in an exactly solvable setting. We then analyze both issues in the three-anyon sector, treating the anyonic phase $\theta=\frac{\pi}{k}$ as a continuous parameter and restricting attention to the linear states. Finally, we extend the analysis to an arbitrary number of anyons. As in the two-anyon case, we find multiple fixed points for $N$-anyon contact interactions. The corresponding spectrum varies continuously with $\theta$ and remains above the unitarity bound.

\subsection{Two Anyons}

Recall that the two-anyon angular momenta are $\ell=2n+\theta/\pi$, where
$n\in\mathbb Z$; see~\eqref{eq:angular-momenta}. Hence, by treating the
anyonic phase as a continuous variable, we can meaningfully extend it to the
range $\theta=\frac{\pi}{k}\in(0,2\pi]$. The corresponding radial equation is
\begin{equation}
  \left[-\frac{1}{2\mu}\frac{\dd^2}{\dd r^2}
  +\frac{\ell^2-\frac14}{2\mu r^2}
  +\frac12\mu r^2
  -{\theta\over\mu}\,\delta^{(2)}(\bm r)\right]u_\ell(r)
  =E_{2n}u_\ell(r)\,,
  \qquad \ell=2n+\theta/\pi\,,
  \label{eq:radial-Schrödinger2p}
\end{equation}
where $\mu=\frac{m}{2}=\frac12$ is the reduced mass. The dimension of the
corresponding operator is $\Delta_{2n}=1+E_{2n}$, where the first term is the
center-of-mass ground-state energy.

For $\ell=\theta/\pi\in[0,1]$, the fixed point with the attractive contact
interaction in~\eqref{eq:radial-Schrödinger2p} selects, from the two
small-$r$ behaviors in~\eqref{smallr}, the more singular solution
$u_\theta(r)=r^{\frac12-{\theta\over\pi}}e^{-\frac14r^2}$. The remaining solutions
in this range are regular and are given by
\begin{equation}\la{chiralsol}
u_\ell(r)=r^{\frac12+|2n+\theta/\pi|}e^{-\frac14r^2}\,,
\qquad
E_{2n}=1+|\ell|=1+|2n+\theta/\pi|\,,
\qquad
n\ne0\,,
\qquad
\theta\in[0,\pi]\,.
\end{equation}
For $\theta\in[0,\pi]$, this is the complete story discussed in the main text.

As $\theta$ reaches $\pi$, however, the norm of the vacuum solution above develops
a logarithmic divergence at the origin. The corresponding operator reaches
the unitarity bound, $\Delta=1$, and behaves as a free field. This is therefore the two-anyon realization of the general normalizability issue that we have discussed in Section~\ref{chiralsec}. At this point, $\theta=\pi$, the $n=-1$ sector enters the range $|2-\theta/\pi|<1$, where both behaviors
$u_{\theta-2\pi}=r^{\frac12\pm(\theta/\pi-2)}e^{-\frac14r^2}$ are normalizable. %
At $\theta>\pi$, the attractive contact interaction selects the more singular of the two. At the same time, the $n=0$ solution leaves the regime $\ell=\theta/\pi<1$, so its
regular solution must be chosen:
$u_\theta(r)=r^{\frac12+\theta/\pi}e^{-\frac14r^2}$. In particular, the
vacuum solution is
\begin{equation}\la{spec}
\Psi_\text{vac}(r)
=e^{\ii\theta\varphi/\pi-\frac14r^2}
\times
\left\{
\begin{array}{llll}
r^{-\theta/\pi}
&E=1-\theta/\pi\,,&n=0\,,
&0\le\theta\le\pi
\\
e^{-2\ii\varphi}\,r^{-(2-\theta/\pi)}
&E=\theta/\pi-1\,,&n=-1\,,
&\pi\le\theta<2\pi
\end{array}
\right.\,.
\end{equation}
The corresponding spectrum is plotted in Figure~\ref{n=2spec}, side by side with the spectrum of the stable fixed point. 
\begin{figure}[t] 
\centering
\includegraphics[width=0.7\textwidth]{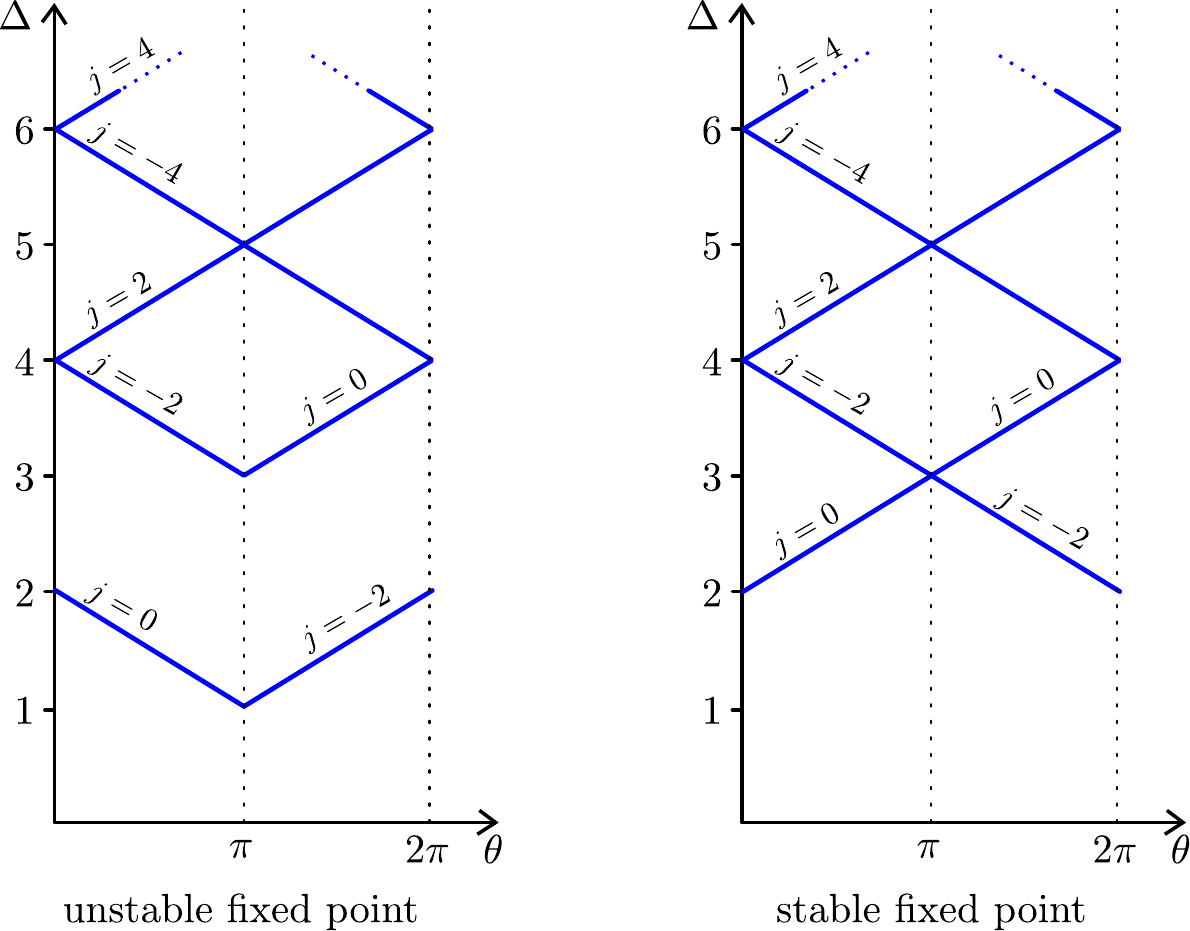} 
\caption{Spectrum of the two-particle sector. Here $j=2n$ is the relative orbital angular momentum.}\la{n=2spec}
\end{figure}
As can be seen in that figure, it is continuous across $\theta=\pi$.

For $\pi<\theta<2\pi$, we have
\begin{equation}
\left.(\Delta_0,\Delta_{-2})\right|_{\pi<\theta<2\pi}
=(2+\theta/\pi,\theta/\pi)\,.
\end{equation}
That is, as $\theta\to\pi$ from below and above, the quantizations of the two states are such that $(\Delta_0,\Delta_{-2})=(1,3)$ and $(\Delta_0,\Delta_{-2})=(3,1)$, respectively. These states carry angular momenta $(J_0,J_{-2})=(1,-1)$, and correspond to the operators
\begin{equation}
    \label{O0Om2}
    \mathcal{O}_0=\Phi^2\,,
    \qquad
    \mathcal{O}_{-2}
    =\Phi\,\partial_{\bar z}^{\,2}\Phi
    -\bigl(\partial_{\bar z}\Phi\bigr)^2\,.
\end{equation}
At $\theta=\pi$, there is a dimension-one operator that behaves as a free field and must be added to the action. Slightly below $\theta=\pi$, this operator is $\mathcal{O}_0$, whereas slightly above $\theta=\pi$, it is $\mathcal{O}_{-2}$. We denote it by $\mathbb{O}$. The composite operator $\mathbb{O}\mathbb{O}^{\dagger}$ has dimension $\Delta=2+2\left|1-\frac{\theta}{\pi}\right|$. When added to the action with the appropriate sign, it triggers an RG flow toward the stable fixed point.

Furthermore, the operators
\[
    \partial_z\mathbb{O}\,
    \partial_{\bar z}\mathbb{O}^{\dagger}
    \qquad\text{and}\qquad
    \partial_{\bar z}\mathbb{O}\,
    \partial_z\mathbb{O}^{\dagger}
\]
have dimension $\Delta=4+2\left|1-\frac{\theta}{\pi}\right|$. They are marginal at $\theta=\pi$ and irrelevant otherwise.\footnote{The operator $\mathbb{O}^2(\mathbb{O}^{\dagger})^2$ is also marginal at $\theta=\pi$, but it affects only the four-particle sector.} Hence, for $\theta>\pi$, there remain two fixed points: one stable and one unstable.

As we will see, an analogous structure appears in the three-anyon problem, although our analysis there is restricted to the subspace of linear states. %

\subsection{Three Anyons}

For three anyons, the center of mass and the total relative scale are defined as
\begin{equation}
 Z=\frac{z_1+z_2+z_3}{3}\,,\qquad
 r^2
 =\sum_{a=1}^3|z_a-Z|^2
 =\frac13\left(
 |z_1-z_2|^2+|z_2-z_3|^2+|z_3-z_1|^2
 \right).
 \label{eq:three-r}
\end{equation}
For the remaining three real variables we use the coordinates
\begin{equation}
 u_1=\frac{z_2-z_3}{\sqrt3\,r}\,,
 \qquad
 u_2=\frac{z_3-z_1}{\sqrt3\,r}\,,
 \qquad
 u_3=\frac{z_1-z_2}{\sqrt3\,r}\,.
 \label{eq:three-u}
\end{equation}
They obey
\begin{equation}
 u_1+u_2+u_3=0\,,\qquad 
 |u_1|^2+|u_2|^2+|u_3|^2=1\,.
 \label{eq:three-normalization}
\end{equation}

In these coordinates, the three-anyon collision limit is $r\to0$ with the $u_i$ fixed. Similarly, two anyons collide when one of the $u_i$'s approaches zero with $r$ fixed.

A common property of all the linear wave functions is their separation of variables form
\begin{equation}
\Psi=\frac{u(r)}{r^{3/2}}\,G(u_1,u_2,u_3)\,e^{-\frac32|Z|^2}\,.
 \label{eq:three-separated-ansatz}
\end{equation}
For such a wave function, the Schr\"odinger equation takes the form\footnote{The eigenvalue of the equation for the $u_i$'s is ${\cal J}({\cal J}-2)$. This equation will not be relevant for us so we will not discuss it here.}
\begin{equation}
 \left[
 -\frac{\dd^2}{\dd r^2}+\frac{({\cal J}-1)^2-\frac14}{r^2}+r^2+g_3\,\delta^{(2)}(r)\right]u(r)
 =2E_{\rm rel}\,u(r)\,,\label{eq:three-u-equation}
\end{equation}
where
\begin{equation}
 {\cal J}=\frac{N(N-1)}{2k}+j=\frac3k+j\label{eq:three-allowed-p}
\end{equation}
is the relative angular momentum. Here, $j=0,\pm2,\pm3,\pm4,\ldots$ is the orbital angular momentum, and $\frac{3}{k}=\frac{3(3-1)}{2k}={3\theta\over\pi}$ is the anyonic statistical contribution~\eqref{exchangephase}.\footnote{Note that there are no linear states with $j=\pm1$. That is because $\sum z_i$ and $\sum\bar z_i$ are components of the center of mass coordinates and therefore do not contribute to the relative angular momentum.}\,\footnote{The $\frac{N}{2k}$ contribution in~\eqref{angmatexact} is an internal angular momentum that does not contribute to the relative one~\eqref{eq:three-allowed-p}.} 
The final term, $g_3\delta^{(2)}(r)$, is new and represents a three-anyon contact interaction. We identify this equation with the two-anyon one in~\eqref{eq:radial-Schrödinger} with
\begin{equation}
\left.\ell\right|_\text{2--anyon}\leftrightarrow\left.({\cal J}-1)\right|_\text{3--anyon}\,.
\end{equation}
The fixed point analysis of $g_3$ is therefore identical to the two-anyon one. Namely, for $|{\cal J}-1|<1$ there are two fixed points with 
\begin{equation}
g_3^\pm=\pm2\pi|{\cal J}-1|\,.
\end{equation}
For $\frac{3}{k}<2$, the relevant solution has $j=0$ orbital angular momentum; for $2<\frac{3}{k}<4$, it has $j=-2$, etc. The short-distance behaviors at the two fixed points are
\begin{equation}
u_\pm(r)\sim r^{\frac12\pm|{\cal J}-1|}\,,\qquad E_\pm=2+2n_r\pm|{\cal J}-1|\,.
\end{equation}
The corresponding linear solutions with $j=0$ are
\begin{equation}
\Psi_{3,-}^{(0)\pm}=\left(r^{\pm|\frac{3}{k}-1|-1}e^{-\frac12r^2}\right)
 (\bar u_1\bar u_2\bar u_3)^{-\frac{1}{k}}\,e^{-\frac32|Z|^2}\,.
 \label{eq:three-singular-gauge}
\end{equation}
Similarly, for $2<\frac{3}{k}<4$ the orbital angular momentum with two possible quantizations is $j=-2$ and the corresponding solutions are
\begin{equation}\label{Psim2}
\Psi_{3,-}^{(-2)\pm}=\Big(\sum_i\bar u_i^2\Big)\left(r^{\pm|\frac{3}{k}-3|-1}e^{-\frac12r^2}\right)
 (\bar u_1\bar u_2\bar u_3)^{-\frac{1}{k}}\,e^{-\frac32|Z|^2}\,.
\end{equation}
Similarly, when $3<\frac{3}{k}<5$, the $j=-3$ angular-momentum sector admits two distinct quantizations. More generally, for every $j\leq -2$, there are two normalizable solutions, or equivalently, two admissible conformal boundary conditions at small $r$ whenever $|j|<\frac{3}{k}<|j|+2$. At each integer value of $\frac{3}{k}\geq 2$, a new operator becomes marginal and must be included in the action. As $\frac{3}{k}$ is increased beyond this value, two fixed points emerge: a repulsive fixed point associated with the more regular solution, at which the operator is irrelevant, and an attractive fixed point associated with the more singular solution, at which the operator is relevant. The spectrum remains continuous and lies above the unitarity bound.

Note that, for every $n < \frac{3}{k} < n+1$, with $3\le n\in\mathbb{Z}$, both states with $j=-n$ and $j=1-n$ admit two possible quantizations. The corresponding solutions satisfy Schrödinger equations with different contact interactions. Therefore, for the unstable fixed points to exist, the coefficient of the contact interaction must be a differential operator whose eigenvalues depend on the orbital angular momentum. This picture is supported by the fact that the operator becoming marginal at $\frac{3}{k}=n$ contains at least two derivatives.

\subsection{More Anyons}

For two anyons, all states are linear, and the preceding analysis is complete.
For more than two anyons, the linear states form only a subset of the
spectrum, while general energy eigenstates are not known. We therefore
restrict the following argument to the linear states, for which the
separation of the total collision radius from the angular variables is
explicit.

Consider the $N$-anyon sector in the limit in which the anyons collide, while no proper subset
of them undergoes an independent collision. As before, we separate the center of mass,
\begin{equation}
 Z_N=\frac{1}{N}\sum_{i=1}^N z_i\,,
\end{equation}
and define the permutation-invariant hyperradius
\begin{equation}
 r^2
 =\sum_{i=1}^N|z_i-Z_N|^2
 =\frac{1}{N}\sum_{i<j}|z_i-z_j|^2
 \longrightarrow0\,.
 \label{eq:N-hyperradius}
\end{equation}
The other shape variables, denoted collectively by $\Omega$, are held fixed and kept away from their lower-particle collision loci. 

A linear state in an angular channel $j$ can be written as
\begin{equation}
 \Psi_{N,j}
 =e^{-\frac{N}{2}|Z_N|^2}\,
 \frac{u_j(r)}{r^{N-3/2}}\,
 G_j(\Omega)\,.
 \label{eq:N-linear-separation}
\end{equation}
Its relative angular momentum is
\begin{equation}
 {\cal J}_j=\frac{N(N-1)}{2k}+j\,,
 \qquad j=0,\pm2,\pm3,\pm4,\ldots\,,
 \label{eq:N-relative-angular-momentum}
\end{equation}
where $j$ is the orbital piece. The separated radial equation is
\begin{equation}
 \left[
 -\frac{\dd^2}{\dd r^2}
 +\frac{\nu_j^2-\frac14}{r^2}
 +r^2
 +g_N\delta_{\rm eff}^{(2)}(r)
 \right]u_j(r)
 =2E_{\rm rel}\,u_j(r)\,,
 \label{eq:N-linear-radial-equation}
\end{equation}
where 
\begin{equation}
 \nu_j
 =
 \left|{\cal J}_j-(N-2)\right|
 =
 \left|
 \frac{N(N-1)}{2k}-(N-2)+j
 \right|\,.
 \label{eq:N-nu-j}
\end{equation}
Here $\delta_{\rm eff}^{(2)}(r)$ is an effective radial pseudopotential. It represents some local $N$-body operator.\footnote{It should not be confused with the literal delta function in the $2(N-1)$-dimensional relative configuration space.}

Near $r=0$, the two independent behaviors are
\begin{equation}
 u_{j,\pm}(r)\sim r^{\frac12\pm\nu_j}\,.
 \label{eq:N-two-radial-powers}
\end{equation}
For $0<\nu_j<1$, both are normalizable and there are two scale-invariant
boundary conditions.  In the normalization inherited from the two-anyon
problem, they correspond to
\begin{equation}
 g_{N,j}^{\pm}=\pm2\pi\nu_j\,.
 \label{eq:N-channel-fixed-points}
\end{equation}
In field theory language, whenever some $\nu_j=0$, a new operator becomes marginal and should be added to the action. This structure supports the existence of a web of unstable fixed points. To prove its existence, it is however necessary to go beyond the linear states, which is beyond the scope of this article.

\section{Giant Axisymmetric Vortices}\label{Giant_vortex_app}

In this appendix, we derive the large-$\lambda$ behavior of the axisymmetric giant vortex states, as quoted in the main text~\eqref{eq:large-ell-crossing} and~\eqref{eq:large-lambda-radial-envelope-energy}.

The energy functional of the axisymmetric giant vortex states is given in~\eqref{eq:radial-functional}. It is convenient to express it in terms of the cumulative charge $Q(r)$ (satisfying $Q'=2\pi r\rho$, $Q(0)=0$ and $Q(\infty)=\lambda$) as
\begin{equation}
    E=E_0+E_{\rm{int}}+E_{\rm{qp}}\,,
\end{equation}
where %
\beq
\frac{E_0}{k}=\frac{1}{2} \int\limits_0^{\infty} \dd r\,Q^{\prime}\left(r^2+\frac{(Q-\ell)^2}{r^2}\right)\,,\qquad
\frac{E_{\rm{int}}}{k}=\int\limits_0^{\infty} \dd r\, \frac{\left(Q^{\prime}\right)^2}{2 r}\,,\qquad
\frac{E_{\rm{qp}}}{k} =\int\limits_0^{\infty} \dd r\, \frac{\left(r Q^{\prime \prime}-Q^{\prime}\right)^2}{8 r^2 Q^{\prime}}\,.
\eeq
A na\"ive scaling analysis suggests that the interaction term $E_{\rm{int}}$ and the quantum pressure term $E_{\rm{qp}}$ are subleading in the large-$\lambda$ limit, so we shall temporarily drop them. Later, we will return to this point and justify this approximation.

To minimize $E_0$, we perform integration by parts on an arbitrary interval $[\epsilon,L]$ and get
\begin{equation}\label{vortexsol}
    \frac{E_0[Q]}{k}=\int\limits_\epsilon^L\dd r\,F(r,Q(r))+\cdots\,,\qquad F(r,Q)=\frac{(Q-\ell)^3}{3 r^3}-r Q\,,
\end{equation}
where $\cdots$ denotes terms independent of variations inside $(\epsilon,L)$. We see that the variational problem reduces to the pointwise minimization of $F$, subject to the constraints $0\leq Q\leq\lambda$ and $Q'\geq0$. The local minimum of $F$ is at $Q_*=\ell+r^2$ (if $Q_*\in[0,\lambda]$), which needs to be compared with the endpoint values at $Q=0,\lambda$ to give the global minimum. We find that minimizing $E_0[Q]$ at fixed $\ell$ and $\lambda$ gives the solution
\begin{equation}\label{eq:giant-vortex-charge}
    Q(r)=
    \begin{dcases}
        0&r<r_0\\
        \ell+r^2&r_0\leq r\leq R\\
        \lambda&R<r
    \end{dcases}\,,\qquad r_0=\sqrt{\ell/2}\,,\qquad R=\sqrt{\lambda-\ell}\,,
\end{equation}
which is consistent with the monotonicity requirement of $Q$. This solution is valid if $r_0\leq R$, or equivalently $\lambda/\ell\geq3/2$, and we will soon see that the favored $\ell$ lies within this regime.\footnote{For $0<\lambda/\ell<3/2$, the minimizer is given by
\begin{equation*}
    Q(r)=
    \begin{dcases}
        0& r<\tilde r_0\\
        \lambda& r\geq\tilde r_0
    \end{dcases}\,,
    \qquad \tilde r_0=\left(\ell^2-\lambda \ell+\frac{\lambda^2}{3}\right)^{1/4}\,,
\end{equation*}
which only has a singular ring with charge $\lambda$ at $\tilde r_0$.} Notably, the cumulative charge profile $Q(r)$ has a jump $\Delta Q(r_0)=3\ell/2$, meaning that there is a singular ring with charge $3\ell/2$ at $r_0$. Outside this ring, there is an annulus with uniform density $\rho_{r_0\leq r\leq R}=1/\pi$. 

The energy of the solution~\eqref{eq:giant-vortex-charge} can be calculated as
\begin{equation}\la{giantvortexE}
    \frac{E_0}{k}=\frac{1}{2}\left[r^2 Q+\frac{(Q-\ell)^3}{3 r^2}\right]_{r_0^-}^{r_0^+}+\frac{1}{2} \int\limits_{r_0}^{R} \dd r\,Q^{\prime}\left[r^2+\frac{(Q-\ell)^2}{r^2}\right]=\frac{\lambda^2}{2}-\lambda\ell+\frac{9\ell^2}{8}\,.
\end{equation}
Equating the energies of the $\ell$ and $\ell+1$ branches gives the crossing points
\begin{equation}
    \lambda_\ell=\frac{9}{8}\left(2\ell+1\right)\,.
\end{equation}
Moreover, treating $\ell$ temporarily as a continuous variable, minimization of $E_0$ over $\ell$ at fixed $\lambda$ gives the favored winding
\begin{equation}
    \ell=\frac{4}{9}\lambda\,,
\end{equation}
or equivalently, $\lambda/\ell\to9/4$ in the large-$\lambda$ limit, which indeed satisfies $\lambda/\ell\geq3/2$. The energy $E_0$ of this favored state is given by
\begin{equation}\label{eq:E_0-min}
    \frac{E_{0,\rm{min}}}{k}=\frac{5}{18}\lambda^2\,,
\end{equation}
which sets a lower bound on the full energy as the two dropped terms $E_{\rm{int}}$ and $E_{\rm{qp}}$ are both positive.

As promised, we now justify the approximation of dropping the terms $E_{\rm{int}}$ and $E_{\rm{qp}}$ in the large-$\lambda$ limit. If we add these terms back in and consider the full energy $E$, the state found above is no longer admissible due to the singular ring and the sharp edge. A nice admissible state would have a smooth ring layer with some radial width $w\ll r_0=O(\lambda^{1/2})$ and charge accumulation $\Delta Q=O(\ell)=O(\lambda)$, an approximately constant bulk with $\rho\approx1/\pi=O(1)$, and a smooth edge layer with some width $u\ll R= O(\lambda^{1/2})$. The contact interaction term thus scales as
\begin{equation}
    \frac{E_{\rm{int}}}{k}=\frac{E_{\rm{int,ring}}}{k}+\frac{E_{\rm{int,bulk}}}{k}+\frac{E_{\rm{int,edge}}}{k}=O\left(\frac{\lambda^{3/2}}{w}\right)+O(\lambda)+O(\lambda^{1/2}u)\,,
\end{equation}
and the quantum pressure term scales as
\begin{equation}
    \frac{E_{\rm{qp}}}{k}=\frac{E_{\rm{qp,ring}}}{k}+\frac{E_{\rm{qp,bulk}}}{k}+\frac{E_{\rm{qp,edge}}}{k}=O\left(\frac{\lambda}{w^2}\right)+O(1)+O\left(\frac{\lambda^{1/2}}{u}\right)\,.
\end{equation}
In addition, the energy $E_0$ changes by at most\footnote{To see this, invert $Q(r)$ and write
\begin{equation*}
    \frac{E_0}{k}=\int\limits_0^\lambda\dd Q\,G(r(Q),Q)\,,\qquad G(r,Q)=\frac{1}{2}\left[r^2+\frac{(Q-\ell)^2}{r^2}\right]\,.
\end{equation*}
Then estimate
\begin{equation}
    \frac{\Delta E_{0,\rm{ring}}}{k}=\int\limits_0^{3 \ell / 2}\dd Q\,[G(r(Q), Q)-G(r_0, Q)]\simeq\int\limits_0^{3 \ell / 2}\dd Q\,\partial_rG(r_0, Q)\delta r(Q)=O(\lambda^{3/2}w)\,,
\end{equation}
and similarly for $\Delta E_{0,\rm{edge}}$.
}
\begin{equation}
    \frac{\Delta E_0}{k}=\frac{\Delta E_{0,\rm{ring}}}{k}+\frac{\Delta E_{0,\rm{edge}}}{k}=O(\lambda^{3/2}w)+O(\lambda u^2).
\end{equation}
Combining all of these contributions and choosing an admissible state -- for example with $w,u=O(1)$, we get an upper bound on the full energy
\begin{equation}\label{Eminoverk}
    \frac{E_{\rm{min}}}{k}\lesssim\frac{5}{18}\lambda^2+O(\lambda^{3/2}).
\end{equation}
Comparing this with the lower bound $E_{\rm{min}}\gtrsim E_{0,\rm{min}}$ in~\eqref{eq:E_0-min}, we conclude that
\begin{equation}
\frac{E_{\rm{min}}}{k}=\frac{5}{18}\lambda^2+o(\lambda^2).
\end{equation}

\subsection{Variational Fission of the Giant Vortex}
\label{app:giant-vortex-fission}

The large-$\lambda$ solution above also gives a simple analytic explanation for the instability of a giant vortex. The essential observation is that the singular core in \eqref{eq:giant-vortex-charge} may be viewed as a localized screened defect embedded in a uniform background of density $1/\pi$. For a defect of winding $s$, the core radius and ring charge are
\begin{equation}
    r_s^2=\frac{s}{2}\,,
    \qquad
    \Delta Q(r_s)=\frac{3s}{2}\,.
\end{equation}
Relative to the uniform background, the empty core removes charge $s/2$, whereas the ring adds charge $3s/2$. The net excess charge of the defect is therefore $s$, equal to its winding. Consequently, the phase winding is screened by the accompanying density excess, and the disturbance of the velocity field vanishes outside the core. To leading order, nonoverlapping defects may thus be treated independently.

This interpretation makes the convex self-energy of a giant vortex manifest. Let
\begin{equation}
    R^2=\lambda-\ell
\end{equation}
be the charge of the uniform background disk of density $1/\pi$. Its contribution to $E_0$ is $R^4/2$. Equation \eqref{giantvortexE} can then be rewritten as
\begin{equation}
    \frac{E_0^{(1)}}{k}
    =
    \frac{R^4}{2}+\frac{5}{8}\ell^2.
    \label{eq:single-giant-defect-energy}
\end{equation}
Thus a screened defect of winding $s$ has self-energy $5s^2/8$. The quadratic dependence on $s$ already suggests that a large defect should fission.

There is also an energy cost for moving a defect away from the center of the trap. In the uniform background,
\begin{equation}
    \boldsymbol v_{\rm bg}(\boldsymbol x)
    =
    -\widehat{\boldsymbol z}\times\boldsymbol x .
\end{equation}
Consider a radially symmetric screened defect of winding $s$, translated from the origin to $\boldsymbol a$. Since its density excess integrates to $s$ and has vanishing dipole moment, translation increases the trapping energy by $s|\boldsymbol a|^2/2$. The constant shift
$\boldsymbol v_{\rm bg}(\boldsymbol x)-\boldsymbol v_{\rm bg}(\boldsymbol x-\boldsymbol a)$
produces an equal increase in the kinetic energy. Hence
\begin{equation}
    \frac{\Delta E_{\rm pos}}{k}
    =
    s|\boldsymbol a|^2.
    \label{eq:defect-position-cost}
\end{equation}

We may now construct an explicit fission trial state. For simplicity, take $\ell$ even and replace the central $\ell$-vortex by two defects of winding
\begin{equation}
    s=\frac{\ell}{2}.
\end{equation}
Each daughter defect has the same local form as \eqref{eq:giant-vortex-charge}, with core radius
\begin{equation}
    r_s^2=\frac{\ell}{4}.
\end{equation}
Place their centers at $\boldsymbol a$ and $-\boldsymbol a$. In the singular limit the two cores are nonoverlapping for $|\boldsymbol a|\geq r_s$; choosing the smallest allowed separation, $|\boldsymbol a|=r_s$, minimizes the positional energy cost. Using \eqref{eq:defect-position-cost}, the leading energy of this configuration is
\beq
    \frac{E_0^{(2)}}{k}
    =
    \frac{R^4}{2}
    +2\left[\frac{5}{8}\left(\frac{\ell}{2}\right)^2\right]
    +2\left(\frac{\ell}{2}\right)\frac{\ell}{4}=
    \frac{R^4}{2}+\frac{9}{16}\ell^2\,.
    \label{eq:split-giant-defect-energy}
\eeq
Comparison with \eqref{eq:single-giant-defect-energy} gives
\begin{equation}
    \frac{E_0^{(2)}-E_0^{(1)}}{k}
    =
    -\frac{\ell^2}{16}\,.
    \label{eq:giant-fission-gain}
\end{equation}
The reduction of the self-energy therefore outweighs the cost of separating the two daughter vortices.

For the preferred axisymmetric branch, $\ell=4\lambda/9+o(\lambda)$, the leading gain is
\begin{equation}
    \frac{E_0^{(2)}-E_0^{(1)}}{k}
    =
    -\frac{\lambda^2}{81}+o(\lambda^2)\,.
\end{equation}
The trial configuration also fits nicely inside the background disk: its outermost core radius is $\sqrt{\ell}$, while
\begin{equation}
    R=\sqrt{\lambda-\ell},
    \qquad
    \frac{\lambda}{\ell}\longrightarrow\frac94>2\,.
\end{equation}
The singular rings can be smoothed and separated by an additional distance of order one. As in the estimates leading to \eqref{Eminoverk}, the resulting interaction, quantum-pressure, and smoothing corrections are at most $O(\lambda^{3/2})$. Thus, for sufficiently large $\lambda$
\begin{equation}
    \frac{E^{(2)}-E^{(1)}}{k}
    \leq
    -\frac{\ell^2}{16}
    +O(\lambda^{3/2})<0\,.
\end{equation}
This proves analytically that the macroscopic axisymmetric giant vortex cannot be the ground state at large $\lambda$.

The same construction applies asymptotically to odd $\ell$ by taking daughter windings $\lfloor\ell/2\rfloor$ and $\lceil\ell/2\rceil$. It can be iterated as long as the daughter windings remain parametrically large. The argument therefore establishes repeated fission of a macroscopic vortex. As we have seen in Section~\ref{splittingsec}, the final preferred state has elementary vortices only.

\section{Crossing Density}
\label{sec:crossing-density}

As discussed in the main text, when \(\lambda\) is varied, the anyon system undergoes a series of first-order phase transitions, producing an increasing number of vortices. The number of vortices is $\sim N/k$ so the creation of vortices accounts for $N/k$ level crossings as we vary $k$ from $k=\infty$ to some finite $k$. 
As we will see in the following, this is only a small fraction of the total number of crossings. 
A parametrically larger number of crossings arises from quantum rotor modes.
The semiclassical saddle breaks rotational invariance, and at finite \(N\) an Anderson tower of rotor modes with angular momentum \(j\) appears.
After quantization, the rotor angular momentum \(j\) is identified with the spin of the wave function and is related to the total angular momentum \(J\) as in Eq.~\eqref{angmatexact},
\begin{equation}
  J=j+\frac{N\lambda}{2}\,.
\end{equation}
The tower is not centered at \(j=0\), but at a value that depends on the configuration of the solid: changing \(\lambda\) shifts this value and, in turn, selects a different angular-momentum state.

In the semiclassical picture, in Coulomb gauge, the saddle has the angular momentum of eq.~\eqref{eq:double-scaling-angular-momentum}:
\begin{equation}
  J_{\rm cl}
  =-k\int \dd^2x \,\rho \,\partial_\varphi\chi
  +\frac{k\lambda^2}{2}
  =N\left(W(\lambda)+\frac{\lambda}{2}\right) .
\end{equation}
We can think of \(W(\lambda)\) as minus the average winding,
\begin{equation}
  W(\lambda)
  =-\frac{1}{\lambda}\int \dd^2x \,\rho \,\partial_\varphi\chi
  =-\frac{\int \dd^2x \,\rho \,\partial_\varphi\chi}
  {\int \dd^2x \, \rho},
\end{equation}
which is naturally of order \(O(N^0)\).

The expression for \(J_{\rm cl}\) is not necessarily quantized, since it is only a classical approximation.
The quantization of the angular momentum is restored by the rotor states.
Consider a slow rotation \(\Omega\) of the solid.
The angular momentum and energy transform as
\begin{align}
  J = J_{\rm cl}+I\Omega+\cdots,\qquad
      E[J] = E_{\rm cl}+\frac{(J-J_{\rm cl})^2}{2I} + \cdots,
\end{align}
where the moment of inertia \(I\) is of order \(N\).
The energetically favored state \(j\) is the one that minimizes
\begin{equation}
  J-J_{\rm cl}
  =j+\frac{N\lambda}{2}-J_{\rm cl}
  =j - N W(\lambda),
\end{equation}
and we denote it by
\begin{equation}
  j_*(\lambda;N)
  =\left\lfloor N W(\lambda)+\frac12\right\rfloor.
\end{equation}
As \(\lambda\) varies, so does \(W(\lambda)\), and every time \(N W(\lambda)\) crosses a half-integer \(n+1/2\), a new quantum state is preferred.
As long as the solid does not undergo a phase transition in which additional vortices are produced, \(W(\lambda)\) varies continuously, and the crossings \(\lambda_n\) are the solutions of
\begin{equation}
  NW (\lambda_n) = n+\frac12,
  \qquad n\in\mathbb Z.
\end{equation}
From this relation we can derive the density of crossings.
At each crossing, \(j_*\) jumps by \(1\), so
\begin{equation}
  \Delta(NW)=1
  \qquad\Longrightarrow\qquad
  N W'(\lambda)\,\Delta\lambda=1,
\end{equation}
and the crossing density is therefore
\begin{equation}
  \frac{\dd n_{\rm cross}}{\dd\lambda}
  =N\left|W'(\lambda)\right|,
\end{equation}
which scales as \(O(N)\) in the double scaling limit.\footnote{The counting is slightly different if the classical solid has \(\mathbb{Z}_s\) symmetry, which can happen at small \(\lambda\) or if, for example, the number of vortices \(N_v\) is a hexagonal number.
  In this case, rotations are identified up to \(2\pi/s\), the angular momenta take the form \(j=j_0+ns\), \(n\in\mathbb{Z}\), and the minimum becomes
  \begin{equation}
    j_*(\lambda;N)
    =j_0+s\left\lfloor\frac{NW-j_0}{s}+\frac12\right\rfloor.
  \end{equation}
  The crossings occur at the values \(\lambda_n\) such that
  \begin{equation}
    NW(\lambda_n)=j_0+s\left(n+\frac12\right),
    \qquad n\in\mathbb Z,
  \end{equation}
  and the crossing density is
  \begin{equation}
    \frac{\dd n_{\rm cross}}{\dd\lambda}
    =\frac{N}{s}\left|W'(\lambda)\right|.
  \end{equation}}

\begin{figure}[t]
  \centering
  \includegraphics[width=\textwidth]{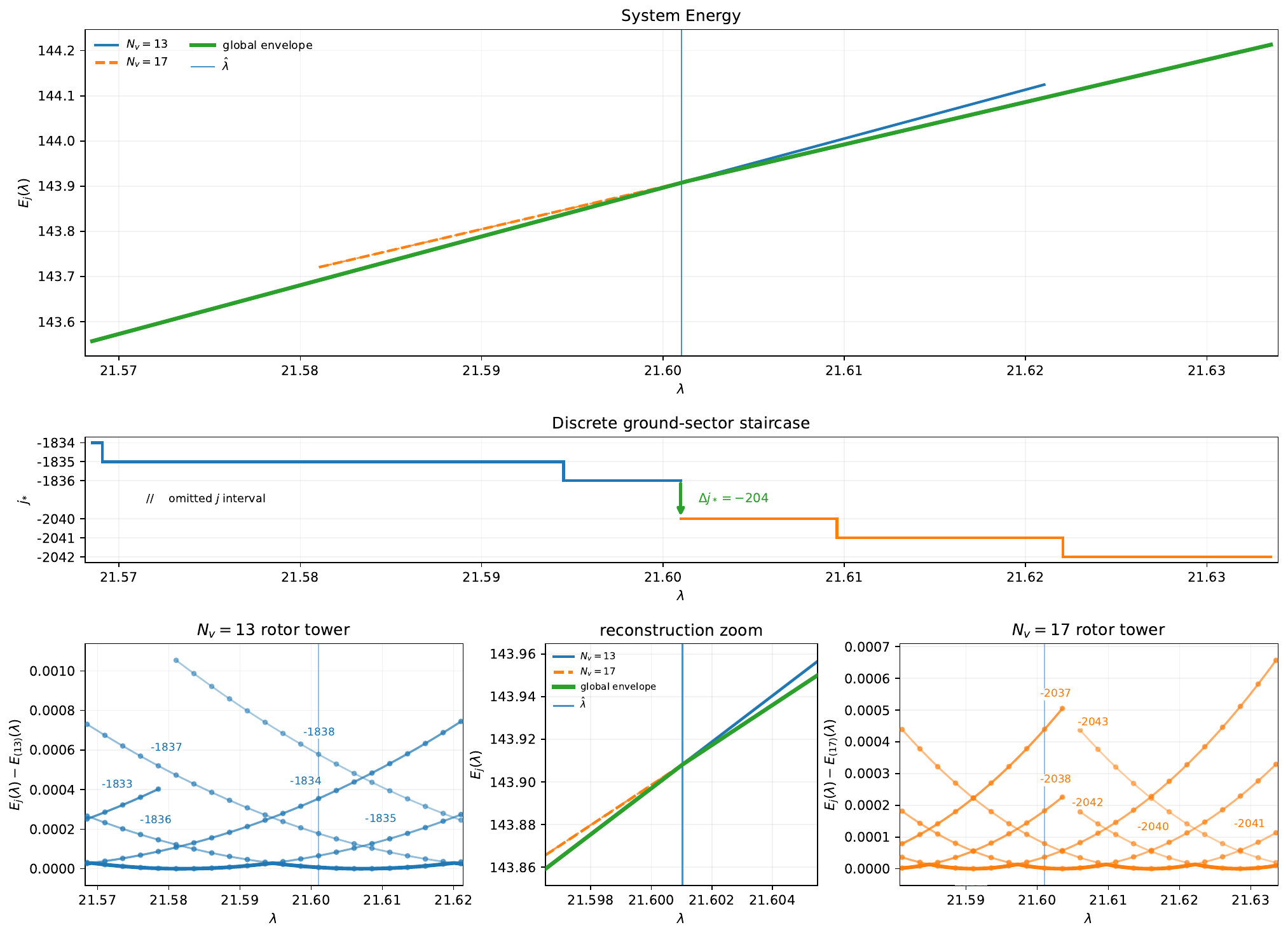}
  \caption{Some crossings around \(\lambda=21\) for a system of \(N=200\) anyons. Around \(\lambda \approx 21.6\), the system undergoes a first-order phase transition from \(N_v=13\) to \(N_v=17\), and the angular momentum \(j_*\) jumps by \(| \Delta j_* | = 204\). Within each phase, adjacent rotor modes cross with \(|\Delta j_*| =1\). %
  }
  \label{fig:around-lambda-20}
\end{figure}

At the values of \(\lambda\) where the ground state changes, \(W(\lambda)\) is discontinuous and \(j_*\) has a single jump of order \(N\).
However, over an interval \((\lambda_1,\lambda_2)\) there will be \(O(\lambda_2-\lambda_1)\) such transitions, and these large jumps do not change the overall parametric estimate of \(O(N)\) for the number of crossings.
All in all, if we vary \(\lambda\) over a finite interval, the total number of crossings takes the form of a sum over the classical saddles of the variation of \(W\) between consecutive phase transitions:
\begin{equation}
  n_{\rm cross}(\lambda_1\to\lambda_2)
  =N\sum_A|\Delta W_A|+\cdots.
\end{equation}
In Figure~\ref{fig:around-lambda-20} we report the result of a numerical computation showing both some \(\Delta j_{*} = 1\) crossings associated with the quantum rotors and a large \(\Delta j_{*}\) crossing due to the creation of new vortices.

For sufficiently large \(\lambda\), where our mesoscopic approximation applies, we can estimate \(W(\lambda)\).
Let \(Q(r)\) and \(Q_v(r)\) denote the total charge and the vortex charge inside a circle of radius \(r\), respectively:
\begin{align}
  Q(r) = 2 \pi \int_0^r r' \dd r' \rho(r'), \qquad Q_v(r) = 2 \pi \int_0^r r' \dd r' \rho_v(r').
\end{align}
By construction, \(\partial_{\varphi}\chi(r)=Q_v(r)\), so after a change of variables the function \(W(\lambda)\) becomes
\begin{equation}
  W(\lambda)
  =-\frac{2 \pi}{\lambda}\int_0^{\infty} r \dd r \, \rho(r) \, \partial_{\varphi}\chi(r)
  =-\frac{1}{\lambda}\int_0^{\lambda}\dd Q \, Q_v(Q).
\end{equation}
Here \(Q_v(Q)\) is the vortex charge inside a circle containing total charge \(Q\).
For perfect screening, which we expect as \(\lambda\to\infty\), \(\rho(r)=\rho_v(r)\) and \(Q_v(Q)=Q\).
It follows that
\begin{equation}
  W(\lambda)=-\frac{\lambda}{2},
\end{equation}
and the crossing density is asymptotically
\begin{equation}
  \frac{\dd n_{\rm cross}}{\dd\lambda}=\frac{N}{2}\,.
\end{equation}
A small refinement of this formula takes into account that, for smaller values of \(\lambda\), screening occurs only in the bulk and there are \(N_v<\lambda\) vortices.
Since the vortices are generally located in the bulk, we can approximate \(Q_v(Q)\) by
\begin{equation}
  Q_v(Q)=
  \begin{cases}
    Q & \text{if $Q<N_v$,} \\
       N_v & \text{if $Q>N_v$,}
  \end{cases}
\end{equation}
which gives the estimate
\begin{equation}
  \frac{\dd n_{\rm cross}}{\dd\lambda}
  =\frac{N}{2} \eta^2.
\end{equation}

\clearpage
\bibliographystyle{JHEP}
\bibliography{Ref}

@article{LeinaasMyrheim,
  author = {J. M. Leinaas and J. Myrheim},
  title = {On the theory of identical particles},
  journal = {Il Nuovo Cimento B Series 11},
  volume = {37},
  number = {1},
  pages = {1--23},
  year = {1977},
  doi = {10.1007/BF02727953},
}

@article{Wilczek1982,
  author = {Frank Wilczek},
  title = {Magnetic Flux, Angular Momentum, and Statistics},
  journal = {Physical Review Letters},
  volume = {48},
  number = {17},
  pages = {1144--1146},
  year = {1982},
  doi = {10.1103/PhysRevLett.48.1144},
}

@article{NishidaSon,
  author = {Yusuke Nishida and Dam T. Son},
  title = {Nonrelativistic conformal field theories},
  journal = {Physical Review D},
  volume = {76},
  pages = {086004},
  year = {2007},
  doi = {10.1103/PhysRevD.76.086004},
  eprint = {0706.3746},
}

@article{XuSC2025,
  author = {Fan Xu and Zheng Sun and Jiayi Li and Ce Zheng and Cheng Xu and Jingjing Gao and Tongtong Jia and Yanfei Su and Kenji Watanabe and Takashi Taniguchi and Bingbing Tong and Li Lu and Jinfeng Jia and Zhiwen Shi and Shengwei Jiang and Junhao Lin and Yuanbo Zhang and Yang Zhang and Shiming Lei and Xiaoxue Liu and Tingxin Li},
  title = {Signatures of unconventional superconductivity near reentrant and fractional quantum anomalous Hall insulators},
  year = {2026},
  eprint = {2504.06972},
}

@article{LiAnyonTrion2026,
  author = {Weijie Li and Christiano Wang Beach and Chaowei Hu and
            Takashi Taniguchi and Kenji Watanabe and Jiun-Haw Chu and
            Ata{\c{c}} Imamo{\u{g}}lu and Ting Cao and Di Xiao and Xiaodong Xu},
  title = {Signatures of fractional charges via anyon--trions in twisted {MoTe$_2$}},
  journal = {Nature},
  volume = {651},
  pages = {48--53},
  year = {2026},
  doi = {10.1038/s41586-026-10101-w},
}

@article{ShiSenthilPRX2025,
  author = {Zhengyan Darius Shi and T. Senthil},
  title = {Doping a fractional quantum anomalous Hall insulator},
  journal = {Physical Review X},
  volume = {15},
  pages = {031069},
  year = {2025},
  doi = {10.1103/kcm5-hx56},
  eprint = {2409.20567},
}

@article{ShiSenthilPNAS2025,
  author = {Zhengyan Darius Shi and T. Senthil},
  title = {Anyon delocalization transitions out of a disordered fractional quantum anomalous Hall insulator},
  journal = {Proceedings of the National Academy of Sciences},
  volume = {122},
  number = {51},
  pages = {e2520608122},
  year = {2025},
  doi = {10.1073/pnas.2520608122},
  eprint = {2506.02128},
}

@article{Senthil2026,
  author = {T. Senthil},
  title = {Fractionalized metals from doped anyons: Application to {tMoTe2}},
  year = {2026},
  eprint = {2607.11484},
}

@article{Laughlin1988,
  author = {R. B. Laughlin},
  title = {Superconducting Ground State of Noninteracting Particles Obeying Fractional Statistics},
  journal = {Physical Review Letters},
  volume = {60},
  number = {25},
  pages = {2677--2680},
  year = {1988},
  doi = {10.1103/PhysRevLett.60.2677},
  note = {Erratum: Phys. Rev. Lett. 61, 379 (1988)},
}

@article{FetterHannaLaughlin1989,
  author = {A. L. Fetter and C. B. Hanna and R. B. Laughlin},
  title = {Random-phase approximation in the fractional-statistics gas},
  journal = {Physical Review B},
  volume = {39},
  number = {13},
  pages = {9679--9681},
  year = {1989},
  doi = {10.1103/PhysRevB.39.9679},
}

@article{Chen1989,
  author = {Yi-Hong Chen and Frank Wilczek and Edward Witten and Bertrand I. Halperin},
  title = {On Anyon Superconductivity},
  journal = {International Journal of Modern Physics B},
  volume = {3},
  pages = {1001--1067},
  year = {1989},
  doi = {10.1142/S0217979289000725},
}

@article{LeeFisher1989,
  author = {Dung-Hai Lee and Matthew P. A. Fisher},
  title = {Anyon superconductivity and the fractional quantum Hall effect},
  journal = {Physical Review Letters},
  volume = {63},
  number = {8},
  pages = {903--906},
  year = {1989},
  doi = {10.1103/PhysRevLett.63.903},
}

@article{WenZee1990,
  author = {X. G. Wen and A. Zee},
  title = {Compressibility and superfluidity in the fractional-statistics liquid},
  journal = {Physical Review B},
  volume = {41},
  number = {1},
  pages = {240--253},
  year = {1990},
  doi = {10.1103/PhysRevB.41.240},
}

@article{BanksLykken1990,
  author = {Tom Banks and Joseph D. Lykken},
  title = {Landau--Ginzburg description of anyonic superconductors},
  journal = {Nuclear Physics B},
  volume = {336},
  number = {3},
  pages = {500--516},
  year = {1990},
  doi = {10.1016/0550-3213(90)90439-K},
}

@article{LykkenSonnenscheinWeiss1990,
  author = {Joseph D. Lykken and Jacob Sonnenschein and Nathan Weiss},
  title = {Anyonic superconductivity},
  journal = {Physical Review D},
  volume = {42},
  number = {6},
  pages = {2161--2165},
  year = {1990},
  doi = {10.1103/PhysRevD.42.2161},
}

@article{DaiEtAl1992,
  author = {Q. Dai and J. L. Levy and A. L. Fetter and C. B. Hanna and R. B. Laughlin},
  title = {Quantum mechanics of the fractional-statistics gas: Random-phase approximation},
  journal = {Physical Review B},
  volume = {46},
  number = {9},
  pages = {5642--5677},
  year = {1992},
  doi = {10.1103/PhysRevB.46.5642},
}

@article{DuMehtaSon2021,
  author = {Yi-Hsien Du and Umang Mehta and Dam Thanh Son},
  title = {Rotons in Anyon Superfluids},
  journal = {JHEP},
  volume = {03},
  pages = {101},
  year = {2021},
  doi = {10.1007/JHEP03(2021)101},
  eprint = {2012.07991},
}

@article{JackiwPiClassicalQuantal,
  author = {R. Jackiw and So-Young Pi},
  title = {Classical and quantal nonrelativistic Chern--Simons theory},
  journal = {Physical Review D},
  volume = {42},
  number = {10},
  pages = {3500--3513},
  year = {1990},
  doi = {10.1103/PhysRevD.42.3500},
  note = {Erratum: Phys. Rev. D 48, 3929 (1993)},
}

@article{BergmanLozano,
  author = {O. Bergman and G. Lozano},
  title = {Aharonov--Bohm Scattering, Contact Interactions, and Scale Invariance},
  journal = {Annals of Physics},
  volume = {229},
  number = {2},
  pages = {416--427},
  year = {1994},
  doi = {10.1006/aphy.1994.1013},
  eprint = {hep-th/9302116},
}

@article{LundholmRougerie2015,
  author = {Douglas Lundholm and Nicolas Rougerie},
  title = {The average field approximation for almost bosonic extended anyons},
  journal = {Journal of Statistical Physics},
  volume = {161},
  pages = {1236--1267},
  year = {2015},
  doi = {10.1007/s10955-015-1382-y},
  eprint = {1505.05982},
}

@article{CorreggiLundholmRougerie2017,
  author = {Michele Correggi and Douglas Lundholm and Nicolas Rougerie},
  title = {Local density approximation for the almost-bosonic anyon gas},
  journal = {Analysis and PDE},
  volume = {10},
  pages = {1169--1200},
  year = {2017},
  doi = {10.2140/apde.2017.10.1169},
  eprint = {1611.00942},
}

@article{LarsonLundholm2018,
  author = {Simon Larson and Douglas Lundholm},
  title = {Exclusion bounds for extended anyons},
  journal = {Archive for Rational Mechanics and Analysis},
  volume = {227},
  pages = {309--365},
  year = {2018},
  doi = {10.1007/s00205-017-1161-9},
  eprint = {1608.04684},
}

@article{CorreggiDuboscqLundholmRougerie2019,
  author = {Michele Correggi and Romain Duboscq and Douglas Lundholm and Nicolas Rougerie},
  title = {Vortex patterns in the almost-bosonic anyon gas},
  journal = {EPL (Europhysics Letters)},
  volume = {126},
  pages = {20005},
  year = {2019},
  doi = {10.1209/0295-5075/126/20005},
  eprint = {1901.10739},
}

@article{Girardot2020,
  author = {Th{\'e}otime Girardot},
  title = {Average Field Approximation for Almost Bosonic Anyons in a Magnetic Field},
  journal = {Journal of Mathematical Physics},
  volume = {61},
  pages = {071901},
  year = {2020},
  doi = {10.1063/1.5143205},
  eprint = {1910.09310},
}

@article{Nguyen2024,
  author = {Dinh-Thi Nguyen},
  title = {2D attractive almost-bosonic anyon gases},
  year = {2024},
  eprint = {2409.00409},
}

@article{GirardotLee2024,
  author = {Th{\'e}otime Girardot and Jinyeop Lee},
  title = {Derivation of the Chern--Simons--Schr{\"o}dinger equation from the dynamics of an almost-bosonic-anyon gas},
  year = {2026},
  eprint = {2412.13080},
}

@article{AtaeiLundholmGirardot2025,
  author = {Alireza Ataei and Douglas Lundholm and Th{\'e}otime Girardot},
  title = {Microscopic derivation of the stationary Chern--Simons--Schr{\"o}dinger equation for almost-bosonic anyons},
  year = {2025},
  eprint = {2504.17488},
}

@article{Visconti2025,
  author = {Fran{\c{c}}ois L. A. Visconti},
  title = {Average-field approximation for dilute almost-bosonic anyons},
  journal = {Calculus of Variations and Partial Differential Equations},
  volume = {65},
  pages = {260},
  year = {2026},
  doi = {10.1007/s00526-026-03428-9},
  eprint = {2507.01102},
}

@article{AtaeiEtAl2025,
  author = {Alireza Ataei and Ask Ellingsen and Filippa Getzner and Th{\'e}otime Girardot and Douglas Lundholm and Dinh-Thi Nguyen},
  title = {Nonlinear Landau levels in the almost-bosonic anyon gas},
  year = {2026},
  eprint = {2510.14679},
}

@article{FavrodOrlandoReffert2018,
  author = {Samuel Favrod and Domenico Orlando and Susanne Reffert},
  title = {The large-charge expansion for Schr{\"o}dinger systems},
  journal = {JHEP},
  volume = {12},
  pages = {052},
  year = {2018},
  doi = {10.1007/JHEP12(2018)052},
  eprint = {1809.06371},
}

@article{KravecPal2019,
  author = {S. M. Kravec and Sridip Pal},
  title = {Nonrelativistic Conformal Field Theories in the Large Charge Sector},
  journal = {JHEP},
  volume = {02},
  pages = {008},
  year = {2019},
  doi = {10.1007/JHEP02(2019)008},
  eprint = {1809.08188},
}

@article{XieHeDasSarma1990,
  author = {X. C. Xie and Song He and S. Das Sarma},
  title = {Finite-size studies of semion systems},
  journal = {Physical Review Letters},
  volume = {65},
  number = {5},
  pages = {649--652},
  year = {1990},
  doi = {10.1103/PhysRevLett.65.649},
}

@article{PomeauRica1994,
  author = {Yves Pomeau and Sergio Rica},
  title = {Dynamics of a Model of Supersolid},
  journal = {Physical Review Letters},
  volume = {72},
  number = {15},
  pages = {2426--2429},
  year = {1994},
  doi = {10.1103/PhysRevLett.72.2426},
}

@article{BoisvertFaddaKulpYazdi,
  author = {Mathieu Boisvert and Shehab Hossam Fadda and Justin Kulp and Ramtin M. Yazdi},
  title = {Revisiting Schr{\"o}dinger CFTs: Factorization, Massless Particles, and a Path to the Bootstrap},
  year = {2026},
  eprint = {2510.26872},
}

@article{JackiwPiSolitons,
  author = {R. Jackiw and So-Young Pi},
  title = {Soliton solutions to the gauged nonlinear Schr{\"o}dinger equation on the plane},
  journal = {Physical Review Letters},
  volume = {64},
  number = {25},
  pages = {2969--2972},
  year = {1990},
  doi = {10.1103/PhysRevLett.64.2969},
}

@article{JackiwPiReview,
  author = {R. Jackiw and So-Young Pi},
  title = {Self-Dual Chern--Simons Solitons},
  journal = {Progress of Theoretical Physics Supplement},
  volume = {107},
  pages = {1--40},
  year = {1992},
  doi = {10.1143/PTPS.107.1},
}

@book{Khare,
  author = {Avinash Khare},
  title = {Fractional Statistics and Quantum Theory},
  edition = {2},
  publisher = {World Scientific},
  address = {Singapore},
  year = {2005},
  isbn = {9789812561602},
  doi = {10.1142/5752},
}

@article{Doroud2016,
  author = {Nima Doroud and David Tong and Carl Turner},
  title = {On superconformal anyons},
  journal = {JHEP},
  volume = {01},
  pages = {138},
  year = {2016},
  doi = {10.1007/JHEP01(2016)138},
  eprint = {1511.01491},
}

@article{Mashkevich1994,
  author = {Stefan Mashkevich and Jan Myrheim and K{\aa}re Olaussen and Ronald Rietman},
  title = {The nature of the three-anyon wave functions},
  journal = {Physics Letters B},
  volume = {348},
  number = {3--4},
  pages = {473--480},
  year = {1995},
  doi = {10.1016/0370-2693(95)00139-C},
  eprint = {hep-th/9412119},
}

@article{Sporre1992,
  author = {M. Sporre and J. J. M. Verbaarschot and I. Zahed},
  title = {Four anyons in a harmonic well},
  journal = {Physical Review B},
  volume = {46},
  number = {9},
  pages = {5738--5741},
  year = {1992},
  doi = {10.1103/PhysRevB.46.5738},
}

@article{Lundholm2017,
  author = {Douglas Lundholm},
  title = {Many-anyon trial states},
  journal = {Physical Review A},
  volume = {96},
  pages = {012116},
  year = {2017},
  doi = {10.1103/PhysRevA.96.012116},
  eprint = {1608.05067},
}

@article{ChitraSen1992,
  author = {R. Chitra and Diptiman Sen},
  title = {Ground state of many anyons in a harmonic potential},
  journal = {Physical Review B},
  volume = {46},
  number = {17},
  pages = {10923--10930},
  year = {1992},
  doi = {10.1103/PhysRevB.46.10923},
}

@article{Dyson:1962es,
  author = {Freeman J. Dyson},
  title = {Statistical Theory of the Energy Levels of Complex Systems. I},
  journal = {Journal of Mathematical Physics},
  volume = {3},
  number = {1},
  pages = {140--156},
  year = {1962},
  doi = {10.1063/1.1703773},
}

@article{Wiegmann:2005eh,
  author = {A. Zabrodin and P. Wiegmann},
  title = {Large N expansion for the 2D Dyson gas},
  journal = {Journal of Physics A: Mathematical and General},
  volume = {39},
  pages = {8933--8964},
  year = {2006},
  doi = {10.1088/0305-4470/39/28/S10},
  eprint = {hep-th/0601009},
}

@article{AkemannByun2019,
  author = {Gernot Akemann and Sung-Soo Byun},
  title = {The high temperature crossover for general 2D Coulomb gases},
  journal = {Journal of Statistical Physics},
  volume = {175},
  pages = {1043--1065},
  year = {2019},
  doi = {10.1007/s10955-019-02276-6},
  eprint = {1808.00319},
}

@article{AGORLargeN,
  author = {Luis Alvarez-Gaum{\'e} and Domenico Orlando and Susanne Reffert},
  title = {Large charge at large N},
  journal = {JHEP},
  volume = {12},
  pages = {142},
  year = {2019},
  doi = {10.1007/JHEP12(2019)142},
  eprint = {1909.02571},
}

@article{HellermanKrichevskiyOrlandoEtAl2024,
  author = {Simeon Hellerman and Daniil Krichevskiy and Domenico Orlando and Vito Pellizzani and Susanne Reffert and Ian Swanson},
  title = {The unitary Fermi gas at large charge and large N},
  journal = {JHEP},
  volume = {05},
  pages = {323},
  year = {2024},
  doi = {10.1007/JHEP05(2024)323},
  eprint = {2311.14793},
}

@article{DondiResurgence,
  author = {Nicola Dondi and Ioannis Kalogerakis and Domenico Orlando and Susanne Reffert},
  title = {Resurgence of the large-charge expansion},
  journal = {JHEP},
  volume = {05},
  pages = {035},
  year = {2021},
  doi = {10.1007/JHEP05(2021)035},
  eprint = {2102.12488},
}

@article{SinghLargeCharge,
  author = {Hersh Singh},
  title = {Large-charge conformal dimensions at the O(N) Wilson--Fisher fixed point},
  year = {2022},
  eprint = {2203.00059},
}

@article{BadelSemiclassics,
  author = {Gil Badel and Gabriel Cuomo and Alexander Monin and Riccardo Rattazzi},
  title = {The epsilon expansion meets semiclassics},
  journal = {JHEP},
  volume = {11},
  pages = {110},
  year = {2019},
  doi = {10.1007/JHEP11(2019)110},
  eprint = {1909.01269},
}

@article{GrassiKomargodskiTizzano,
  author = {Alba Grassi and Zohar Komargodski and Luigi Tizzano},
  title = {Extremal correlators and random matrix theory},
  journal = {JHEP},
  volume = {04},
  pages = {214},
  year = {2021},
  doi = {10.1007/JHEP04(2021)214},
  eprint = {1908.10306},
}

@article{HellermanOrlandoSQCD,
  author = {Simeon Hellerman and Domenico Orlando},
  title = {Large R-charge EFT correlators in N=2 SQCD},
  year = {2021},
  eprint = {2103.05642},
}

@article{HellermanExponential,
  author = {Simeon Hellerman},
  title = {On the exponentially small corrections to N=2 superconformal correlators at large R-charge},
  year = {2021},
  eprint = {2103.09312},
}

@article{Babadi2013,
  author = {Mehrtash Babadi and Brian Skinner and Michael M. Fogler and Eugene Demler},
  title = {Universal behavior of repulsive two-dimensional fermions in the vicinity of the quantum freezing point},
  journal = {EPL (Europhysics Letters)},
  volume = {103},
  pages = {16002},
  year = {2013},
  doi = {10.1209/0295-5075/103/16002},
  eprint = {1212.1493},
}

@article{Astrakharchik2021,
  author = {G. E. Astrakharchik and I. L. Kurbakov and D. V. Sychev and A. K. Fedorov and Yu. E. Lozovik},
  title = {Quantum phase transition of a two-dimensional quadrupolar system},
  journal = {Physical Review B},
  volume = {103},
  pages = {L140101},
  year = {2021},
  doi = {10.1103/PhysRevB.103.L140101},
  eprint = {2012.14008},
}

@article{BruunNelson2014,
  author = {Georg M. Bruun and David R. Nelson},
  title = {Quantum hexatic order in two-dimensional dipolar and charged fluids},
  journal = {Physical Review B},
  volume = {89},
  pages = {094112},
  year = {2014},
  doi = {10.1103/PhysRevB.89.094112},
  eprint = {1401.2237},
}

@article{BedanovGadiyakLozovik1985Lindemann,
  author = {V. M. Bedanov and G. V. Gadiyak and Yu. E. Lozovik},
  title = {On a modified Lindemann-like criterion for 2D melting},
  journal = {Physics Letters A},
  volume = {109},
  number = {6},
  pages = {289--291},
  year = {1985},
  doi = {10.1016/0375-9601(85)90617-6},
}

@article{BedanovGadiyakLozovik1985Melting,
  author = {V. M. Bedanov and G. V. Gadiyak and Yu. E. Lozovik},
  title = {Melting of two-dimensional crystals},
  journal = {Soviet Physics JETP},
  volume = {61},
  number = {5},
  pages = {967--973},
  year = {1985},
  note = {Russian original: Zh. Eksp. Teor. Fiz. 88, 1622--1633 (1985)},
}

@article{Khrapak2020,
  author = {Sergey A. Khrapak},
  title = {Lindemann melting criterion in two dimensions},
  journal = {Physical Review Research},
  volume = {2},
  pages = {012040},
  year = {2020},
  doi = {10.1103/PhysRevResearch.2.012040},
  eprint = {2002.00651},
}

@article{Levitt2026,
  author = {Antoine Levitt and Douglas Lundholm and Nicolas Rougerie},
  title = {Magnetic Thomas--Fermi theory for 2D abelian anyons},
  journal = {SciPost Physics Core},
  volume = {9},
  number = {1},
  pages = {018},
  year = {2026},
  doi = {10.21468/SciPostPhysCore.9.1.018},
  eprint = {2504.13481},
}

@article{AmelinoBak,
  author = {Giovanni Amelino-Camelia and Dongsu Bak},
  title = {Schr{\"o}dinger self-adjoint extension and quantum field theory},
  journal = {Physics Letters B},
  volume = {343},
  number = {1-4},
  pages = {231--238},
  year = {1995},
  doi = {10.1016/0370-2693(94)01448-L},
  eprint = {hep-th/9406213},
}

@article{Sen1992,
  author = {Diptiman Sen},
  title = {Some supersymmetric features in the spectrum of anyons in a harmonic potential},
  journal = {Physical Review D},
  volume = {46},
  number = {4},
  pages = {1846--1852},
  year = {1992},
  doi = {10.1103/PhysRevD.46.1846},
}

@article{Doroud2018,
  author = {Nima Doroud and David Tong and Carl Turner},
  title = {The conformal spectrum of non-Abelian anyons},
  journal = {SciPost Physics},
  volume = {4},
  number = {4},
  pages = {022},
  year = {2018},
  doi = {10.21468/SciPostPhys.4.4.022},
  eprint = {1611.05848},
}

@article{GirardotRougerie2021,
  author = {Th{\'e}otime Girardot and Nicolas Rougerie},
  title = {Semiclassical limit for almost fermionic anyons},
  journal = {Communications in Mathematical Physics},
  volume = {387},
  pages = {427--480},
  year = {2021},
  doi = {10.1007/s00220-021-04164-1},
  eprint = {2101.04457},
}

@article{JackiwPiTimeDependent,
  author = {R. Jackiw and So-Young Pi},
  title = {Time-Dependent Chern--Simons Solitons and Their Quantization},
  journal = {Physical Review D},
  volume = {44},
  number = {8},
  pages = {2524--2532},
  year = {1991},
  doi = {10.1103/PhysRevD.44.2524},
}

@article{Feshbach:1958nx,
  author = {Herman Feshbach},
  title = {Unified theory of nuclear reactions},
  journal = {Annals of Physics},
  volume = {5},
  number = {4},
  pages = {357--390},
  year = {1958},
  doi = {10.1016/0003-4916(58)90007-1},
}

@article{ByrdLuNocedalZhu,
  author = {Richard H. Byrd and Peihuang Lu and Jorge Nocedal and Ciyou Zhu},
  title = {A Limited Memory Algorithm for Bound Constrained Optimization},
  journal = {SIAM Journal on Scientific Computing},
  volume = {16},
  number = {5},
  pages = {1190--1208},
  year = {1995},
  doi = {10.1137/0916069},
}

\end{document}